# Human-Hardware-in-the-Loop simulations for systemic resilience assessment in cyber-socio-technical systems

Francesco Simone[1]*, Marco Bortolini[2], Giovanni Mazzuto[3], Giulio di Gravio[1], Riccardo Patriarca[1]

[1]Sapienza University of Rome, Department of Mechanical and Aerospace Engineering, Via Eudossiana 18, 00184 Rome (RM), Italy

[2]Alma Mater Studiorum – University of Bologna, Department of Industrial Engineering, Viale del Risorgimento 2, 40136, Bologna (BO), Italy

[3]Università Politecnica delle Marche, Department of Industrial Engineering and Mathematical Sciences, Via Brecce Bianche, 60100, Ancona (AN), Italy

*Corresponding author: francesco.simone@uniroma1.it

**Abstract:** Modern industrial systems require updated approaches to safety management, as the tight interplay between cyber-physical, human, and organizational factors has driven their processes toward increasing complexity. In addition to dealing with known risks, managing system resilience acquires great value to address complex behaviors pragmatically. This manuscript starts from the System-Theoretic Accident Model and Processes (STAMP) as a modelling initiative for such complexity. The STAMP can be natively integrated with simulation-based approaches, which however fail to realistically represent human behaviors and their influence on the system performance. To overcome this limitation, this paper proposes a Human-Hardware-in-the-Loop (HHIL) modeling and simulation framework aimed at supporting a more realistic and comprehensive assessments of systemic resilience. The approach is tested on an experimental oil and gas plant experiencing cyber-attacks, where two personas of operators (experts and novices) work. This research provides a mean to quantitatively assess how variations in operator behavior impact the overall system performance, offering insights into how resilience should be understood and implemented in complex socio-technical systems at large.

## 1. Introduction

Managing safety in industrial systems has become an increasingly challenging task over time. If on one hand, a growing awareness on safety-related themes has provided researchers and practitioners with a wide range of tools and methodologies to identify and to address operational risks; on the other, the continuous technological advancement has made industrial systems progressively more complex. Over the years, industrial processes have undergone a significant transformation. They evolved from predominantly manual, craft-like operations, where workers used machines as tools, to highly automated configurations in which intelligent machines can now perform more and more complicated tasks. This shift has marked a blurred transition from a simple coexistence between humans and machines to scenarios of close collaboration, where both entities work together to enable the system to achieve its goals (Dobra & Dhir, 2020). In addition, the relevance of organizational factors related to such operational processes has been acknowledged (Vredenburgh, 2002). As such, over the last two decades, scholars increasingly emphasized the importance of adopting a systemic perspective for managing safety, suggesting the integration of human, technical, and organizational dimensions alike (Hollnagel et al., 2013; Leveson, 2012). As a result, safety management has become closely tied to the concept of socio-technical systems (STSs) (Baxter & Sommerville, 2011), i.e., systems made of people, machines, and organizations, all tightly and dynamically interacting to ensure that the system fulfills its intended purpose.

The mentioned evolution of industrial systems pointed towards a growing digitalization of STSs. The “technical” dimension evolved into two interconnected facets: (i) a physical part, made up of tangible equipment, and (ii) a digital part, involving all the data and the information exchanged within the cyberspace. This distinctions becomes tangible as cyber-physical systems (CPSs) (Rajkumar et al., 2010) have become increasingly prevalent in industrial environments: while their adoption enhances performance by optimizing processes and enabling better coordination among activities, they also introduce a range of new risks. Notably, digital failures can now result in tangible, real-world physical consequences (Yaacoub et al., 2020). On this premises, the concept of STS has been recently expanded to highlight the additional complexities introduced by the adoption of CPSs: Patriarca et al. (2021) introduced the notion of cyber-socio-technical systems (CSTS), emphasizing how the dual nature of the technical dimension increases also the complexity related to the “social” dimension of STSs, i.e., the human and organizational aspects.

## 1.1. The System-Theoretic Accident Model and Processes (STAMP) and its related techniques

In this complex scenario, among the various systemic models and methods, the System-Theoretic Accident Model and Processes (STAMP) and its related techniques (Leveson, 2012) stood out. The STAMP has been extensively employed not only in scientific research but also in numerous industrial case studies, demonstrating both a strong theoretical foundations, and the practical effectiveness of the resulting analyses (Patriarca et al., 2022).

From a theoretical standpoint, the STAMP combines principles from both systems theory and control theory to offer a novel perspective on systems safety. Specifically, the adoption of a system thinking viewpoint embedded in STAMP incorporates the effects on, and the mutual dependencies among the system’s elements. Meanwhile, drawing from control theory principles, the STAMP views the system as a collection of interrelated elements that must be maintained in a dynamic equilibrium through feedback and control loops. According to the STAMP, whatever safety issue arises when such control mechanisms result inadequate or fail. The whole rationale behind the STAMP and its related techniques builds on this premises: the existence of unsafe (or unsecure) control actions (UCAs). Indeed, analyzing a STAMP model turns out in identifying which control actions may result (or resulted) inadequate and how, to eventually define the safety constraints to ensure this will not happen (again). The System-Theoretic Process Analysis (STPA) and the System-Theoretic Process Analysis for Security (STPA-Sec) formalize this process in a proactive manner, while the Causal Analysis based on Systems Theory (CAST) applies a similar approach, but retrospectively.

Although Patriarca et al. (2022) provides a comprehensive review of the usage of the STAMP and its related techniques, it is here important to emphasize the dynamicity of the field. Indeed, the latest (i.e., 2022, onwards) advancements in STAMP-related research majorly highlight the qualitative nature of the methods as a key limitation, particularly when attempting to systematically and objectively apply their findings in practice. Researchers converge around the idea that, while the STAMP offers a rich, qualitative basis for the mapping of unsafe control actions and systemic causal factors, pairing it with quantitative tools brings out its full potential. Indeed, major research effort has been put in finding ways to associate the STAMP with quantification approaches: research over the past few years has increasingly explored methods to introduce quantitative rigor into STAMP-based safety analysis, highlighting both the advantages and the constraints of these practices. For instance, Ebrahimi et al. (2024) integrated STAMP with fuzzy logic to quantify the interrelationships among control levels, eventually producing interpretable "cause-effect" weights that extend the STAMP beyond its qualitative bounds. However, the quantitative results remained largely dependent on expert judgements and static assessments of interdependencies, without fully freeing the analysis from subjectivity. Similarly, Sun et al. (2024) demonstrated that combining the STAMP with a cascading failure propagation model could enable dynamic risk profiling in real time. The approach was effectively used to quantify how faults propagate across a system, yet it required assumptions about the propagation mechanisms to consider. Additionally, the STAMP and the STPA have been combined with event-tree analysis (Zhu et al., 2025), and Bayesian networks (Nakhal Akel et al., 2025) to estimate the probabilities of accidents occurrences, leveraging probabilistic inference to produce interpretable risk scores. Also in this case, the validity of the results was sensitive to the quality of expert-elicited inputs, and it did not fully capture the system dynamics over time. A similar example can be found in the research by Liao et al. (2025), where STPA was paired with Bayesian networks and Success Likelihood Index Method (SLIM) to model human error likelihood, but, again, results relied on expert-estimated probabilities and validation data. In contrast, approaches that couple STAMP-related techniques with simulations show particular advantages when it comes to the objectivity and scalability of results, permitting to iterate and test over large sets of parameters settings. Examples are the STPA-SPN framework by Bensaci et al. (2023) which used Petri nets and Monte Carlo simulation to directly estimate collision frequencies in multi-robot systems, or the STPA-Sec/S method by Simone et al. (2023)

which enabled the resilience assessment through direct modeling of system dynamics. These simulation-based methods produced empirical distributions of outcomes under varied scenarios and operational conditions, capturing cascading effects, and time dependencies that static inference approaches cannot. While expert judgment remains essential in defining the safety control structure (SCS) and the UCAs, pairing these models with simulations allows for far richer quantitative insights. Indeed, the reviewed literature indicates that while probabilistic and static modeling techniques offer valuable results, integrating STAMP with simulation techniques provides a particularly promising path forward (Nakhal Akel et al., 2024).

On these premises, this paper builds up on simulations approaches presenting a method to enhance the quantitative analysis of STAMP results, while preserving its systemic take on CSTSs.

### 1.2. Objective of this research

Although simulations represent a compelling solution to quantify and explore the results of a STAMP analysis, especially in complex and dynamic environments, two main challenges must be acknowledged (Davis & Marcus, 2016). A first difficulty in simulating CSTSs lies in the inherent simplification that modeling entails. Models are, by design, abstractions of reality and they cannot capture the full complexity of real-world systems (Box, 1976). This limitation may be particularly problematic when dealing with CSTSs and cyber failures that can propagate non-linearly, and in context-dependent ways, often with emergent and unpredictable outcomes, eventually affecting the accuracy and interpretability of obtained results. A second challenge relates to the fact that in CSTSs the presence of human agents has an active role in the system. However, unlike technical components', the human behavior is highly variable and influenced by cognitive, social, and contextual factors that are difficult to formalize into a model.

Building on these premises, this paper addresses the following research question: *how to quantitatively assess the results of a STAMP-based analysis for a CSTS leveraging Human-Hardware-in-the-Loop (HHIL) simulations ?*

With HHIL, we here refer to a simulation approach that integrates both cyber-physical components of the system (i.e., hardware) and real humans interacting with those components. By including human agents directly in the simulation loop, this approach aims to overcome the aforementioned limitations, especially in those scenarios where human decision-making is tightly intertwined with the management of cyber failures. Indeed, this research primarily focuses on CSTSs' cyber disruptions, and the proposed method draws inspiration from – and extends – the STPA-Sec/S approach introduced by Simone et al. (2023). The method is instantiated and tested through a case study involving an experimental plant that reproduces an oil and gas facility. The experimental setup allowed for a controlled observation of the system behaviors and human responses in the presence of simulated cyber-attacks, permitting to argue the effectiveness of the HHIL approach in quantifying the STPA-Sec/S results.

## 2. Method

This section outlines the approach used in this study. It begins with a brief overview of the STPA-Sec/S method, which serves as the foundation for this research. Subsequently, the necessary adaptations to the simulation model are discussed in order to enable the HHIL simulations.

### 2.1. System-Theoretic Process Analysis for Security with Simulations (STPA-Sec/S)

The STPA-Sec/S represents a methodological development of the traditional STPA and its cybersecurity-oriented extension, i.e., STPA-Sec. By combining the system-theoretic foundations of STPA-Sec with simulation-based modeling, STPA-Sec/S introduces a way to move beyond pure qualitative reasoning, and it enables getting some quantitative insights into how complex systems dynamically behaves in response to cyber threats. . This type of results become particularly relevant when analyzing a CSTS cyber resilience, i.e., its ability to anticipate, withstand, recover from, and adapt to cyber disruptions (Björck et al., 2015). Following Simone et al., (2023), the STPA-Sec/S is structured into seven key phases:

- *Step 1. Define the scope of the analysis*. It involves the identification of the system boundaries, its mission, the unacceptable losses and their related hazards, and the associated safety and security constraints. Such elements guide the whole analysis keeping the processes detailed in all the other steps within a specific scope;
- *Step 2. Model the safety control structure*. The system hierarchical SCS is developed outlining the relationships and the interactions, –among controllers, actuators, sensors, and controlled processes. The SCS is then used to guide the digital model development;
- *Step 3. Identify the unsafe and unsecure control actions*. All the control actions within the SCS are analyzed to determine under which conditions they might become unsafe or unsecure. Specifically, a control action may turn inadequate and identify an UCA when: (i) not providing a feedback or a control action leads to an hazard; (ii) providing a feedback or a control action leads to an hazard; (iii) providing a feedback or a control action too early, too late, in wrong order, or with an inappropriate application leads to an hazard; (iv) a feedback or a control action provided are stopped too late, or too soon, leading to an hazard. The identified UCAs are linked to specific functional roles within the system, representing potential vectors for cyber-induced failure;
- *Step 4. Develop the simulation model*. The SCS (obtained in Step 2) is used as a baseline to implement a digital model in a simulation environment. The simulation model includes the representations of the system elements in terms of state variables capable of depicting the system dynamics of interest (following Step 1);
- *Step 5. Define the resilience metrics*. In order to assess the impact of cyber-attacks, one or a set of indicators for the resilience performance shall be identified. These shall be directly linked to the scope of the analysis (Step 1), and they can include physical measures (e.g., flow rate, temperature), cyber metrics (e.g., packet loss rate, latency, jitter) service-level indicators (e.g., production output, availability), or mission-specific outcomes (e.g., compliance with safety thresholds);

- *Step 6. Model the faults and their effects*. Attack scenarios are developed based on the previously identified UCAs in Step 3. The UCAs are translated into possibilities of stressing changes within the digital model during simulations;
- *Step 7. Evaluate the system resilience performance*. The final step involves executing simulation runs under different attack scenarios and comparing system behavior to its nominal state. The difference in performance, as captured by the resilience metrics, allows analysts to compute resilience indices and evaluate the system's ability to cope with the disruptions.

Regarding this final step, the simulation runs can extend beyond simple deterministic input-output computations, instead leveraging systematically stochastic simulations to gain a systemic understanding of the CSTS.

## 2.2. Toward Human-Hardware-in-the-Loop (HHIL) simulations

To increase the realism of the simulation executed in the final phase of STPA-Sec/S (i.e., Step 7), this research integrates HHIL logics. The HHIL simulation architecture is directly grounded in the system's SCS, as defined in the Step 4 of the STPA-Sec/S. However, in this extended use, the SCS supports both the identification of key state variables (i.e., those affected by control actions or shared through feedback), and it serves to map the different modes of interaction between human and technological agents to be captured in the HHIL simulation model. In this context, three distinct types of human-technology interaction can be identified from a simple human-automation SCS schema (cf. **Figure 1**):

- *Automated process control (human monitoring)*. In this configuration, the human operator is excluded from directly controlling the physical process, and they can only monitor the process parameters. This occurs whenever the operator chooses to not to exercise their control capabilities, but instead delegating all the process control to the automated systems;

- *Human indirect process control (human empowering)*. In this case, the human operator influences the process by modifying the logic of the automated control system. Although they are actively involved, the operator's actions are still mediated by the automation, and they do not directly interact with the physical process;
- *Human direct process control (human exclusivity)*. Here, the human operator exerts manual control over the process by directly interacting with physical actuators and reading measurements from the equipment sensors, without relying on any automation intermediaries.

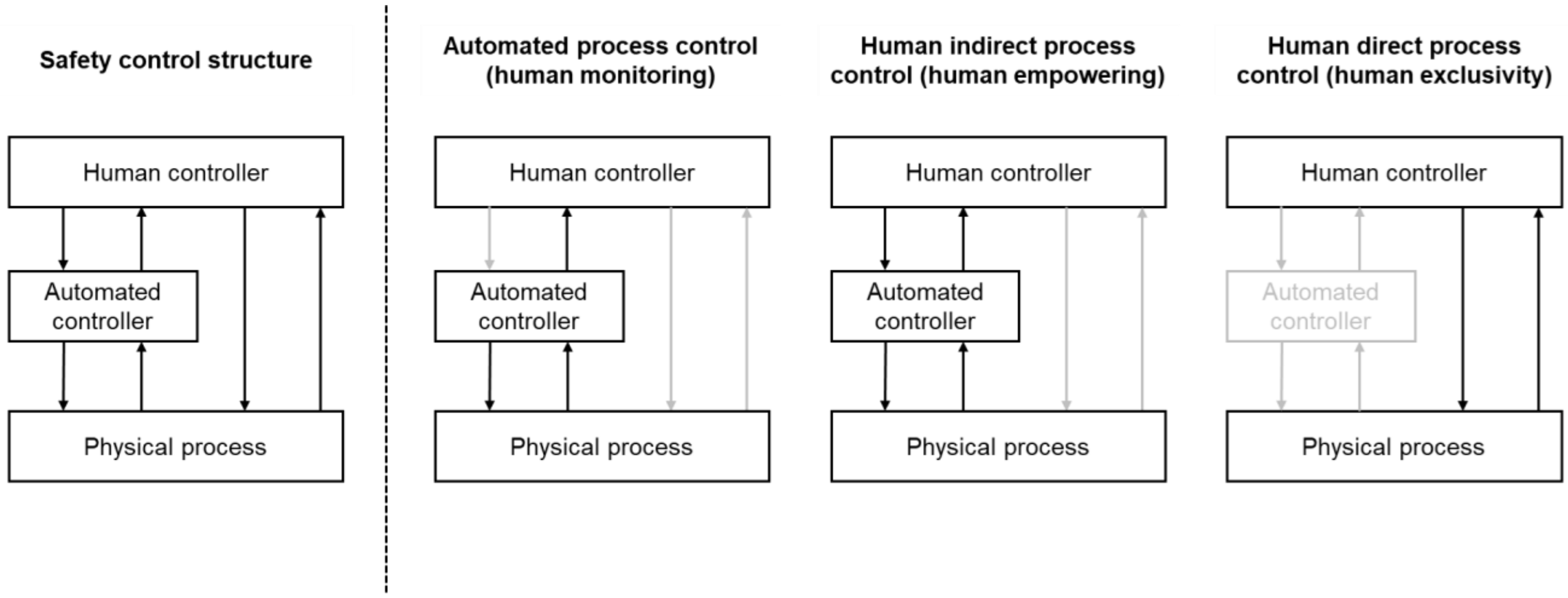

**Figure 1.** An exemplary SCS of a CSTS (left), and the three types of human-technology interactions that may occur during the system operations (right).

The three modes of interaction can occur multiple times and in varying sequences throughout the process operations. Thus, the objective is to develop a simulation model architecture that can dynamically adapt to capture each of these interaction modes whenever they arise (Simone et al., 2024). Based on this premise, the HHIL simulation model incorporates all three modes through an architectural design such as the one illustrated in **Figure 2**. The architecture includes a physical process (i.e., "Physical process" block, cf. **Figure 2**) which must be tangibly reproduced in the real world, for example through a testbed or a mock-up simulator. The existence of a real physical process is critical to enabling the interactions with the human operator (i.e., "Human operator" block, cf. **Figure 2**). Since the system under study is a CSTS, some level of automation must also be incorporated. Accordingly, the architecture also considers sensors (i.e., “Sensor” block, cf. **Figure 2**) and actuators (i.e., “Actuator” block, cf. **Figure 2**). The sensors collect and transmit process data, while the actuators act on the system to modify its physical parameters. Indeed, the data acquired by the sensors are processed by an automated controller (i.e., “Automated controller” block, cf. **Figure 2**), which issues control commands through the actuators. This automated controller – or more typically, a set of controllers – operates under the supervision of a higher-level control system (e.g., a SCADA system) that aggregates information from multiple sources within the process, then formulating appropriate control strategies (i.e., “Data center” block, cf. **Figure 2**). Finally, the model architecture includes a digital counterpart of the physical process (i.e., “Digital model of the system” block, cf. **Figure 2**), which is meant to enable the simulation of scenarios that cannot be safely or practically reproduced in the physical setup, such as destructive or hazardous events.

To support the simulation of all the previously discussed human-technology interaction modes (illustrated in **Figure 1**), the HHIL model architecture must enable specific interfaces between its elements. These shall allow the model to dynamically represent various operational configurations depending on the scenario being simulated. In particular, **Figure 2** outlines the required interaction links. Starting from the physical layer, continuous data exchange must be ensured between the physical process, sensors, actuators, and the automated controller (i.e., solid light blue arrows, cf. **Figure 2**). However, these interactions with the real, tangible process must remain configurable in the model: they must be available, but selectively disabled in simulations involving potentially destructive scenarios. In such cases, the digital model of the system substitutes the real physical process, and the automated controller interacts with this digital representation instead, creating a Hardware-in-the-Loop configuration (i.e., dashed black arrows, cf. **Figure 2**). Conversely, other interfaces between technological components must remain active at all times (i.e., solid black arrows, cf. **Figure 2**). Specifically, the automated controller must continuously communicate with the data center to ensure up-to-date process information, whether the data originate from the physical system or the digital model. The data center, in turn, exchanges this information with the digital model to maintain consistency and enable accurate process simulation when needed. The digital model must also incorporate information related to the human controller. Human-related data can be collected via (e.g.) motion capture systems, cameras, or wearable sensors, and they should be fed into the digital model in a Human-in-the-Loop fashion to enable the simulation of human behavior and its effects on the system performance. To this end, the human controller must be able to interact with the various technological elements in the system, including: sensors (e.g., by directly reading data to monitor process conditions), actuators (e.g., by manually adjusting process parameters), and the automation system (e.g., by modifying control logic via the data center). All these human-technology interfaces are represented by solid orange arrows in **Figure 2**.

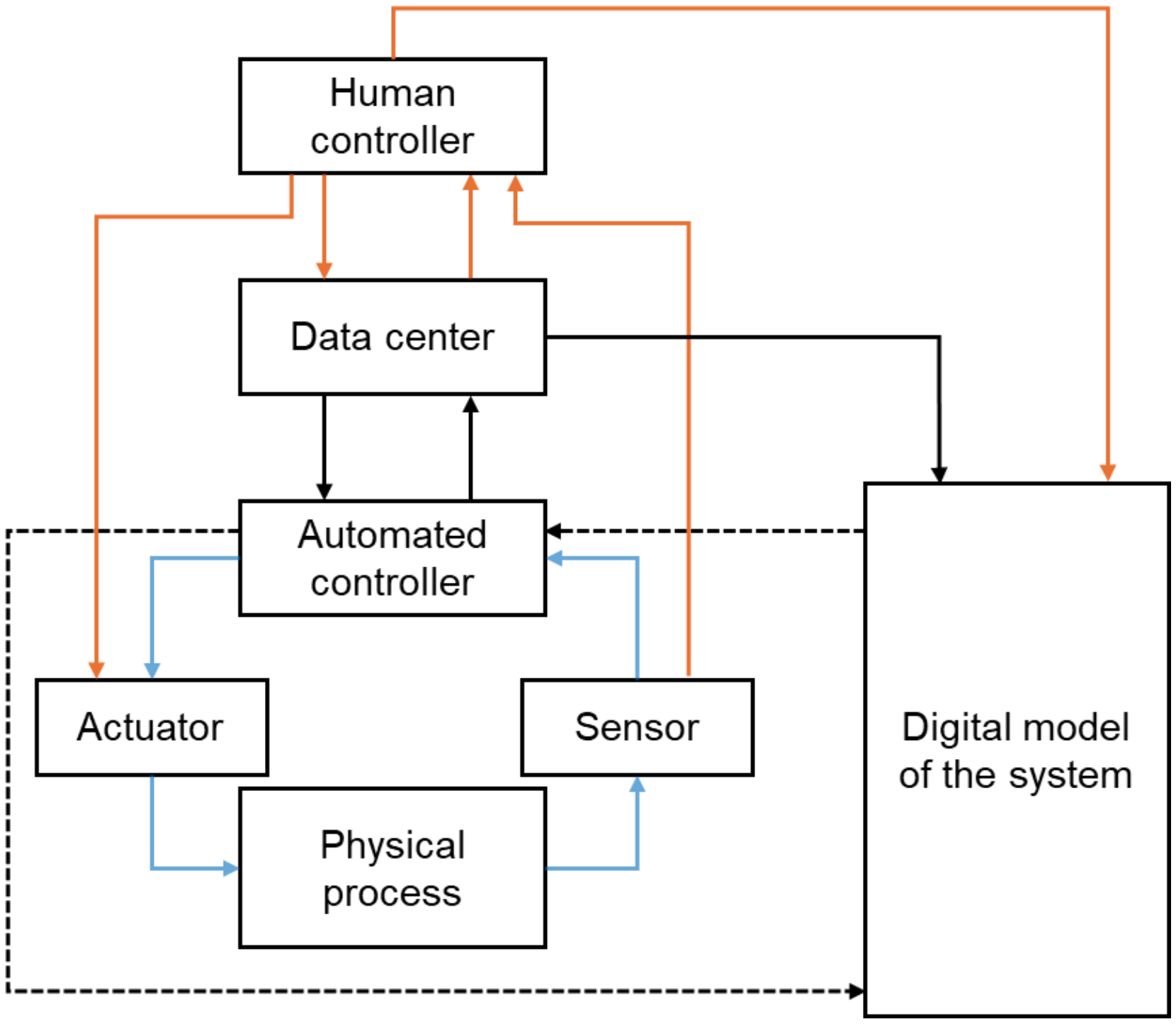

**Figure 2.** HHIL simulation model architecture. Solid black arrows represent interaction links between technological elements that are always active. Dashed black arrows indicate interaction links between technological elements that are permanently available but are used only during the simulation of destructive scenarios. Solid orange arrows represent interaction links between human operators and technological elements that are always active. Solid light blue arrows denote interaction links between technological components that must not be active during destructive scenarios. Adapted from (Simone et al., 2024).

## 3. Case study

This section presents the case study used to test the HHIL modeling and simulation approach proposed in this paper. The analyzed process is firstly introduced, followed by the application of the STPA-Sec/S methodological steps. This latter has been intentionally centered on the development and use of the HHIL model, rather than providing a comprehensive overview of the entire methodology application. Consequently, several details have been intentionally omitted to maintain the focus on the role and the implementation of HHIL simulations within the broader approach.

### 3.1. Process overview

The case study used to instantiate the proposed HHIL extension of STPA-Sec/S is an experimental mock-up plant located in the Department of Industrial Engineering and Mathematical Sciences at the Università Politecnica delle Marche in Ancona, Italy (Di Carlo et al., 2021). A plant scheme is reported in **Figure 3**. The plant is designed to emulate the artificial extraction of natural gas from a depleted well by exploiting the pressure from an adjacent, active, oil well. This approach is commonly adopted in the energy sector to reduce the cost and the complexity of installing mechanical pumps at reservoir depths. In accordance with the “do no significant harm” principles (European Commission, 2021), the plant substitutes crude oil and natural gas with room-temperature water and air, while preserving the structural and functional behavior of an industrial two-phase flow extraction system. The core component of the process is a gas-liquid ejector, which uses a high-pressure water stream to create a vacuum, allowing ambient air to be drawn into the system. This mimics the operation of real-world ejectors used in the oil and gas industry for the transport of a two-phase gas-liquid mixtures. The process begins with a pump drawing water from an open tank, whose pressure is regulated by a solenoid valve (i.e., AV1), before being delivered to the ejector. Once the high-pressure water enters the ejector, it accelerates, converting pressure into kinetic energy to generate a vacuum. This latter draws in air via an inlet pipeline dedicated to air, eventually forming a two-phase air-water mixture at the ejector outlet. The mixture flows into a vertical tank which serves as a phase separator, isolating the gas and the liquid components. The tank outlet are managed through two additional solenoid valves (i.e., AV2 and AV3) which regulate the exit of the water and the air, respectively, allowing for the active control over the liquid level and the tank internal pressure. To automatically monitor and control the plant, a Revolution Pi (RevPi) Core 3 industrial controller is employed. The RevPi collects real-time data from the plant's sensors (see **Table 1** for the complete list of sensors and actuators equipped on the plant), processes it via a PID logic, and commands the actuators (i.e., AV1, AV2, AV3) accordingly. Data and commands are transmitted using a Message Queuing Telemetry Transport (MQTT) publish-subscribe communication architecture, which connects the RevPi to a human-machine interface implemented on both a desktop and a mobile app. The user interface is designed for both real-time supervision of the plant operations, and for enabling the operators to intervene in process control. It provides both live and historical data visualizations for key performance indicators, while offering manual control functionalities for adjusting parameters and tuning the control logics. In this context, the operator plays a critical role in ensuring the whole system reliability, particularly during anomalous or unexpected conditions that go beyond their passive monitoring. If necessary, the operator can physically inspect the plant to gather additional information or perform direct manual interventions, thereby ensuring that the ejector all the other components continue to operate within their intended performance specifications.

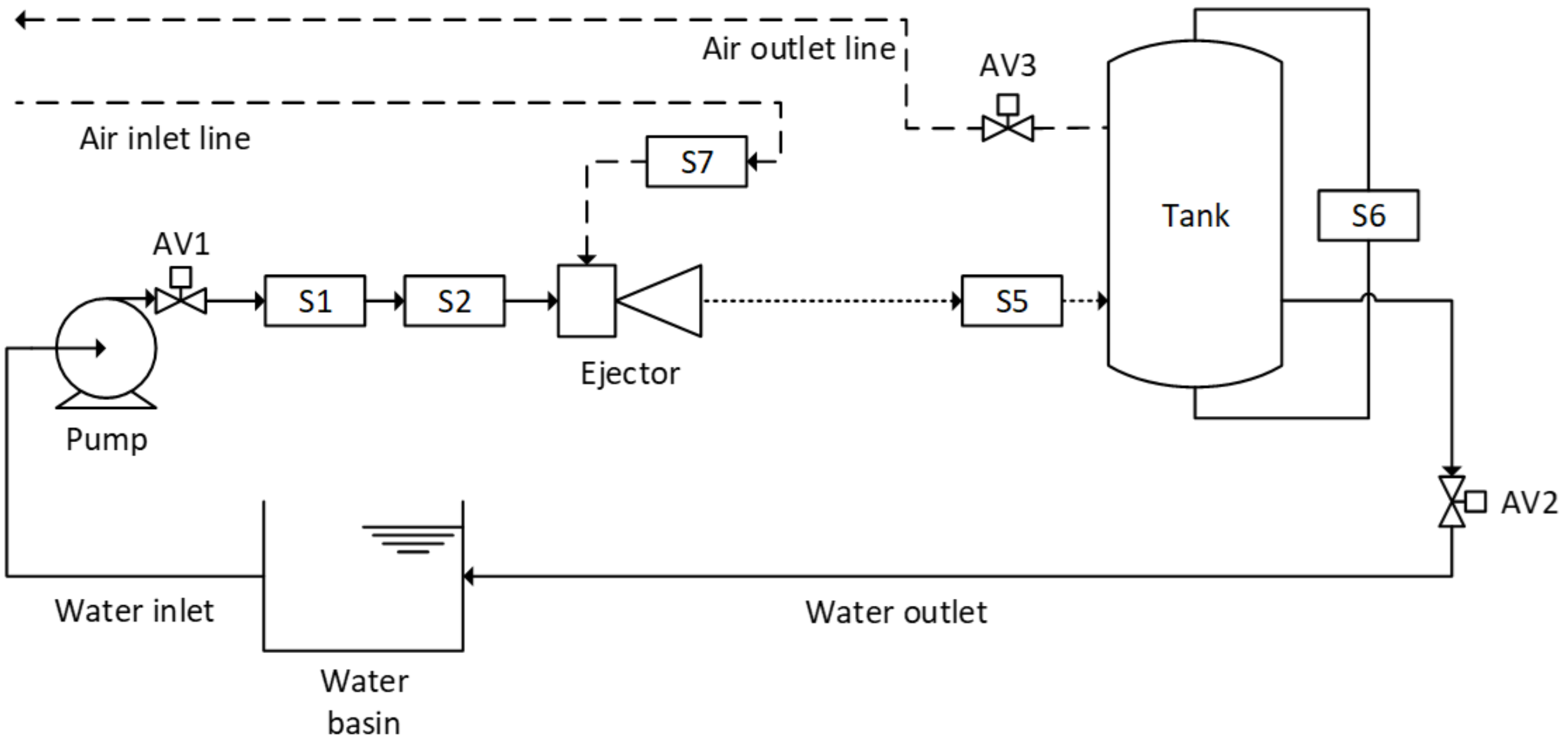

**Figure 3**. Experimental plant scheme. Solid lines represent water flows, dashed lines represent air flows, dotted lines represent the water-air mixture flows.

**Table 1.** List of plant components (sensors and actuators) with corresponding description.

| Component name | Description |
|---|---|
| Sensor S1 | Sensor measuring the pressure of water entering the ejector |
| Sensor S2 | Sensor measuring the flow rate of water entering the ejector |
| Sensor S5 | Sensor measuring the pressure inside the tank |
| Sensor S6 | Sensor measuring the level of water inside the tank |
| Sensor S7 | Sensor measuring the flow rate of air drawn into the ejector |
| Actuator AV1 | Valve regulating the passage of inlet water, it is used to regulate the pressure of water entering the ejector |
| Actuator AV2 | Valve regulating the passage of outlet water, it is used to regulate the level of water inside the tank |
| Actuator AV3 | Valve regulating the passage of outlet air, it is used to regulate the pressure inside the tank |

## 3.2. Applying the STPA-Sec/S steps

The close collaboration between the human operators and the automated technological components in achieving the system's operational objectives makes the above descripted experimental plant a representative example of a CSTS to be used for testing the STPA-Sec/S HHIL approach. Key details of this implementation are reported in the following paragraphs.

### 3.2.1. Define the scope of the analysis

During the problem identification phase (i.e., Step 1, cf. **Section 2.1**), the analysis focused on the potential cyber-induced failures the plant might face. In particular, the focus was posed on those cyber-attacks leading to the violation of critical pressure thresholds, or resulting in insufficient water or air flowrates, in turn compromising the ability to perform the extraction process. The full list of losses, their associated hazards, and the system-level constraints is considered out of scope, as acknowledged early in **Section 3**.

### 3.2.2. Model the safety control structure

Regarding the modeling of the SCS (i.e., Step 2, cf. **Section 2.1**), the resulting SCS is shown in **Figure 4**, where it is possible to notice all the key elements previously described in **Section 3.1** and summarized in **Table 1**.

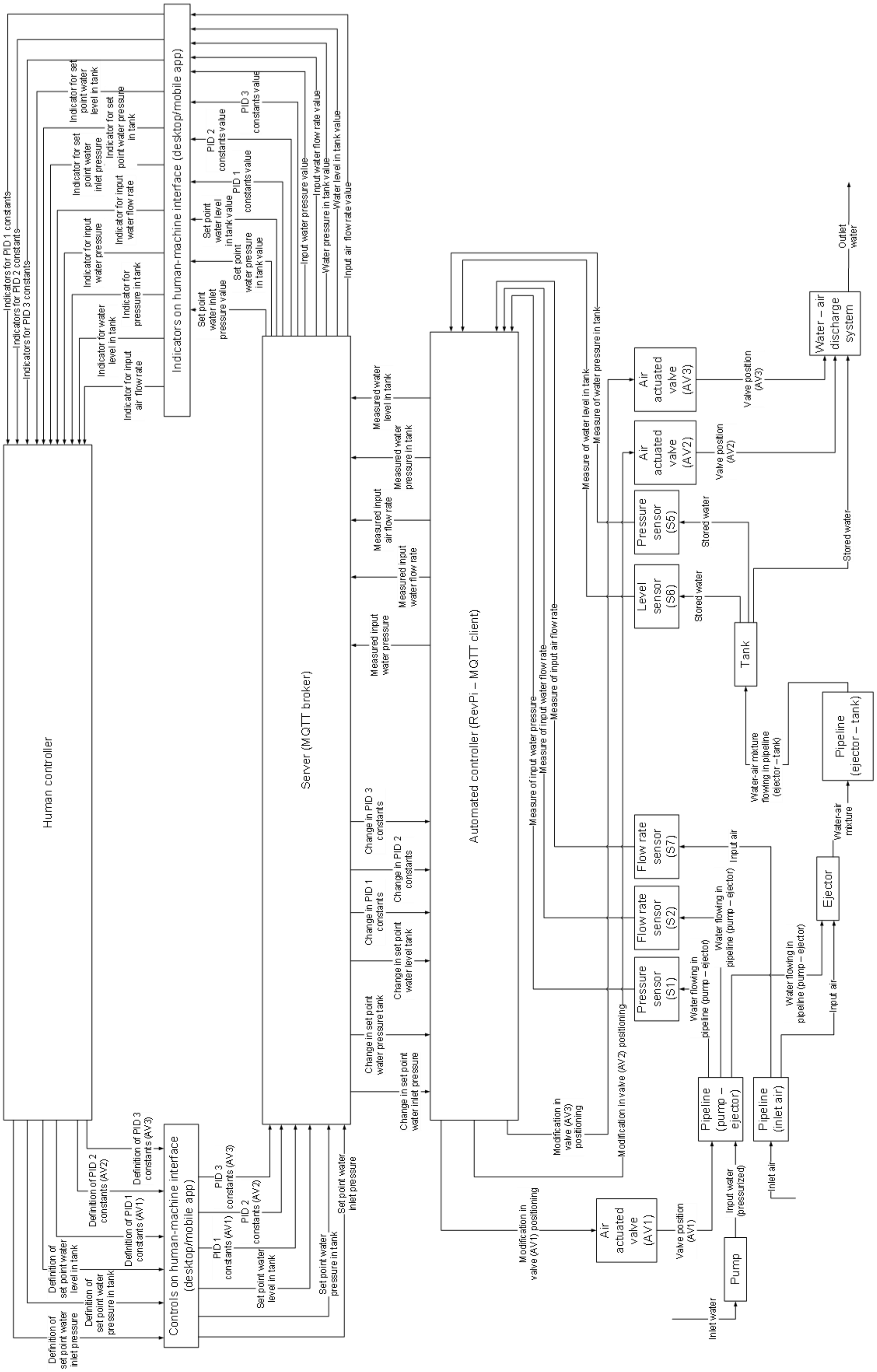

**Figure 4.** SCS of the analyzed CSTS.

### 3.2.3. Identify the unsafe and unsecure control actions

The SCS in **Figure 4** was analyzed to retrieve the system's UCAs. For demonstration purposes, the analysis presented in this paper, and the obtained results focus only on the following UCAs:

- *UCA 1*. The RevPi provides a wrong modification of position for the air actuated valve (AV3) while controlling the amount of air in tank during steady state operations;
- *UCA 2*. The RevPi does not provide any modification in the positioning of the air actuated valve (AV1) while controlling the flow of water in pipeline (pump – ejector) during steady state operations;
- *UCA 3*. The RevPi provides a wrong modification of position for the air actuated valve (AV2) while controlling the water level in tank during steady state operations.

Additional details with respect to the simulation scenarios developed based on these UCAs are later reported in **Section 3.2.5**.

### 3.2.4. Develop the simulation model

In relation to the HHIL model architecture shown in **Figure 2**, the implementation of the blocks labeled "Physical Process", "Sensor", "Actuator", "Automated Controller", and "Data Center", along with their respective interfaces, should be straightforward to interpret, as they directly correspond to the physical components and connections that constitute the mock-up plant described in **Section 3.1**. Greater attention, however, should be devoted to explaining how the remaining two blocks (i.e., the "Human Operator", and the "Digital Model of the System") and their interfaces with the plant's components have been integrated into the simulation model.

The integration of the human operator into the HHIL simulation model has been realized through the development of a human digital twin using the Perception Neuron Studio motion capture system by Noitom. This solution was mainly chosen for its portability and for its ability to perform high-precision motion tracking without the need for external cameras or fixed infrastructure, thus avoiding any modifications to the existing mock-up plant layout. The setup mimicked the one by Bortolini et al. (2025), and it included 17 inertial measurement sensors placed on key joints of the operator's body to capture the full kinematics of their movements at up to 240 frames per second. However, for the purposes of this study, only the pelvis sensor was used to approximate the operator's center of mass, tracking their movements within the mock-up plant environment at a resolution of 90 fps. The raw output of the motion capture system consisted of a time series of three-dimensional coordinates for the pelvis sensor, referenced to a coordinate system anchored to a neutral (i.e., empty) fixed point. Starting from this spatial data, the operator's positions were contextualized into actions within the system, by introducing a set of control volumes, i.e., virtual three-dimensional zones positioned around key components of the plant, such as (e.g.) inspection points. Each control volume was defined as a non-overlapping parallelepiped, with specific origin coordinates and dimensions along the three axes. Accordingly, the operator's actions became an interaction with a given control volume, determined through a presence condition checking whether the pelvis sensor's coordinates fell within the control volume boundaries during a given time frame. Indeed, the processed output of the motion capture system included a table reporting: (i) the "active" control volume, and (ii) the persistence of the operator within it.

On the other hand, the digitalization of the physical process, intended to allow a digital twin to replace the physical plant as the automated controller's target during destructive simulation scenarios, was implemented using the Python-based model developed by Mazzuto et al. (2025). Such digital model estimates the plant process variable by replicating the coupled behavior of two main plant subsystems: the pump-ejector line, and the vertical separation tank. Accordingly, the pump-ejector model simulates how the water pressure drives the air suction, computing pressure values based on the water flow conditions, and enabling the simulated adjustment of the water inlet valve (i.e., AV1). In parallel, the vertical tank model computes the tank internal pressure and the water level over time, using this information to simulate the regulation of the two outlet valves positions (i.e., AV2, and AV3). The two subsystems are dynamically interconnected by a feedback loop since variations in the tank pressure and in the water level must affect the ejector suction performance, too. The model is capable of returning a time series of values for all the simulated plant's components (i.e., sensors and actuators) which precisely integrate with the real ones coming from physical components. As a result, the physical simulation produced a table in which each column corresponded to the state of a specific plant component (as listed in **Table 1**), with the rows capturing how those states evolved over time.

### 3.2.5. Define the resilience metrics

A set of resilience metrics was defined to track the differences between the nominal performance and the disrupted one by means of the eight process parameters of interest (i.e., the one described by the S1, S2, S5, S6, and S7 measures, and by the AV1, AV2, AV3 behavior). Each resilience metric followed a similar structure, specifically:

$$R_i = \frac{As_i - |As_i - Ad_i|}{As_i} \tag{1}$$

where $As_i$ represents the area under the nominal performance curve, $Ad_i$ represents the area under the disrupted performance curve, and $i \in$ [S1, S2, S5, S6, S7, AV1, AV2, AV3] identifies the eight different resilience indicators being defined.

The metrics were based on the well-established "area under the curve" approach (Hosseini et al., 2016), which is conceptually related to the measurement of resilience through the so-called "resilience triangle" (Tierney & Bruneau, 2007).

**Figure 5** illustrates an exemplary $i$-th system performance using an hypothetical time series generated by the plant's digital model. Both the nominal (i.e., dotted line, cf. **Figure 5**) and disrupted (i.e., solid line, cf. **Figure 5**) performance curves are shown, with the resilience indicator $R_i$ represented by the shaded area. At first glance, the definition of the resilience indicators may appear disconnected from the operator's behavior. However, in relation to **Figure 5**, it is important to identify five distinct phases in the disrupted curve that eventually indicate when and how the operator contributes to the overall system performance. Specifically:

- <u>*Steady-state*</u>, cf. **Figure 5** [$t_0$, $t_1$). In a first phase of operations, the system behave under a steady-state condition (i.e., $P_s$, cf. **Figure 5**), with no cyber-attacks nor disruption occurring. As the intended analysis assumes that at some time a cyber-attack will succeed, this phase offers no opportunity for the operator's intervention. Therefore, operator-related data are not relevant in this phase assuming that neither positive nor detrimental effects can be put in place by them;

- *Under attack*, cf. **Figure 5** [$t_1$, $t_2$). At a certain point in time (i.e., $t_1$, cf. **Figure 5**), a cyber-attack starts affecting the plant performance. During this phase, the operator's role refers first to detect the anomaly. The operator's detection time (i.e., $t_2$, cf. **Figure 5**), directly influences the length of this phase: the sooner the operator identifies the attack, the shorter the "under attack" phase lasts;

- *Shutdown*, cf. **Figure 5** [$t_2$, $t_3$). Once the attack is detected, the operator lets the plant enter in the shutdown phase, where the performance declines to a "zero" condition (i.e., $P_0$, cf. **Figure 5**). Although the duration of this phase is not directly governed by operator's behavior, it is indirectly affected by how quickly the operator reacted in the previous phase. Specifically, the time required to reach $P_0$ depends on the extent to which the system was impacted by the attack (i.e., $P_d$, cf. **Figure 5**). For instance, if the performance curve in **Figure 5** represents the water level in the tank (i.e., S6), the earlier the attack is detected, the less the water level rises, making it possible to empty the tank quicker;

- *Plant off*, cf. **Figure 5** [$t_3$, $t_4$). After the shutdown, the plant remains off, entering an unintended yet stable period. During this time, the operator assesses the attack's effect on the system, investigates potential malfunctions on components, and carries out the necessary recalibrations and diagnostics tasks. The length of this phase is directly linked to the duration of the operator's intervention: the quicker these tasks are completed, the sooner the plant can resume its operations;

- *Start-up*, cf. **Figure 5** [$t_4$, ∞). Once all the recovery and verification activities are completed, the operator begins the start-up process (i.e., from $t_4$ onward, cf. **Figure 5**), and the plant transitions back to its original steady-state (i.e., $P_s$, cf. **Figure 5**). The transition can be non-linear and discontinuous, as mimicked by the line in **Figure 5**. Since this phase is largely automated, the operator's behavior has no influence over its duration and, thus, the operator-related data are not included in this phase either.

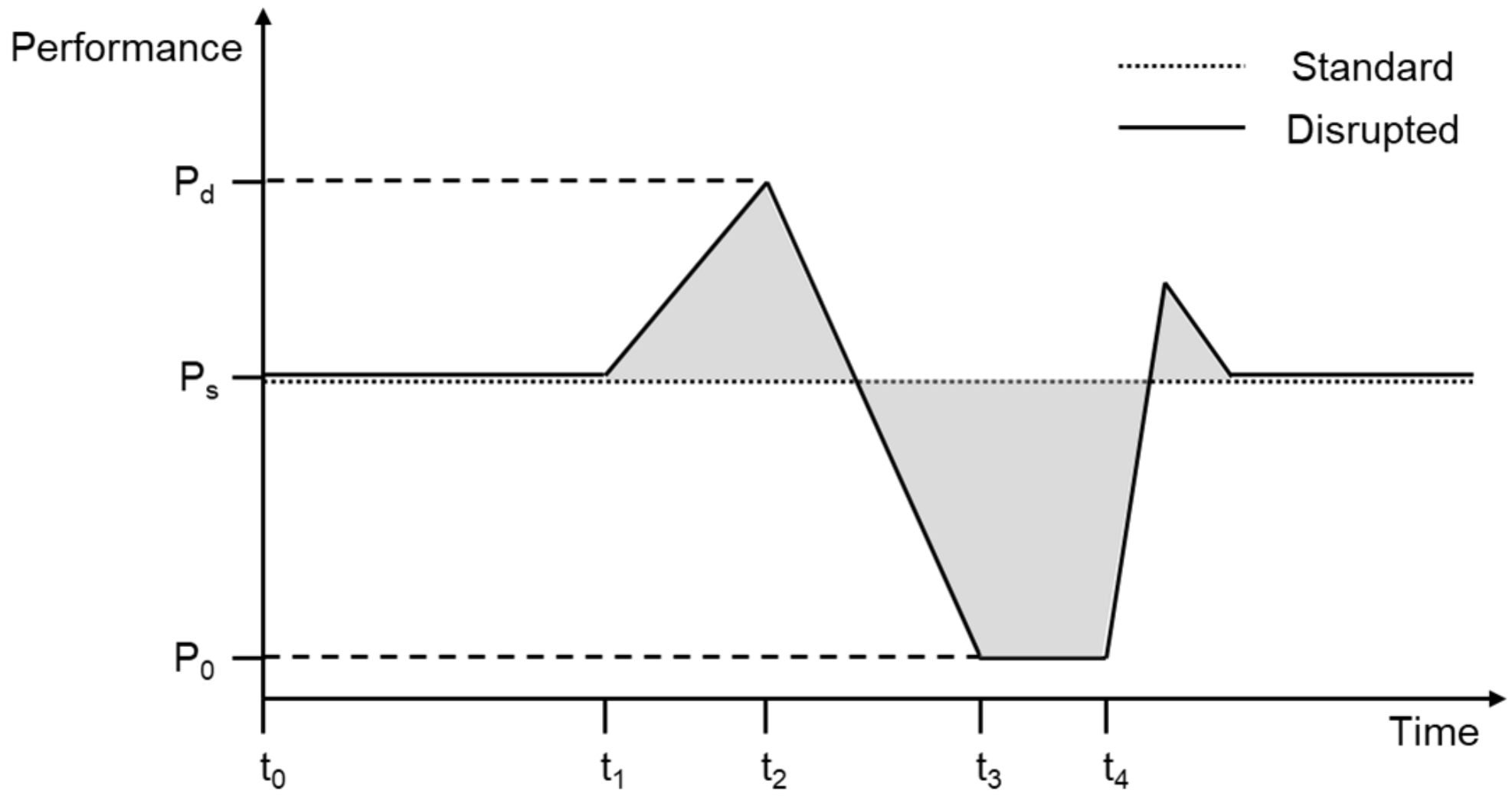

**Figure 5**. Exemplary standard and disrupted performance curves used for calculating the resilience indicator, with key time moments and corresponding performance values.

### 3.2.6. Model the faults and their effects

Based on the selected UCAs listed in **Section 3.2.3**, two representative scenarios were formalized to serve as the foundation for the subsequent resilience analysis. Among all the loss scenarios derived from the STPA-Sec analysis, these two were specifically highlighted by the plant operators as both relevant and feasible for simulation. As such, they were selected and here presented to evaluate the HHIL simulation approach. The scenarios are described below, detailing the faults and their technical implications, as well as the expected responses from the human operator when performing the HHIL simulations:

- *Scenario 1: false low reading of tank level*. The feedback "Measured water level in tank" or "Measure of water level in tank" (cf. **Figure 4**) indicates low level of water inside the tank. However, this reading is incorrect, and it result from a cyber-attack as (e.g.) a stealthy false data injection (FDI) attack that deliberately manipulates sensor readings, or a covert Man-in-the-Middle (MitM) attack that prevents sensor updates while simultaneously injecting malicious code or manipulating control setpoints to alter plant behavior without detection. The anomalous situation prompts the automated controller to change the way AV2 is managed (i.e., UCA 3): assuming the tank to be underfilled, AV2 is closed to avoid additional water to draw out. However, the actual water level continues to rise, and as the water accumulates, the pressure in the tank increases. In this case, the correct provision of the feedback “Measured water pressure in tank” and “Measure of water pressure in tank” make the automated controller aware of the increase in pressure and it reacts by managing AV3 properly, eventually releasing the air excess to maintain a stable pressure level. However, the amount of water produced at the plant’s output decreases due to the closing of AV2, and the rising water level could lead to the tank overflow. Upon detecting the sensor discrepancy and recognizing the associated overflood risk, the operator intervenes by disabling the automated controller to stop any further unintended control action. This manual intervention is aimed at bringing the plant to a stable state. Subsequently, the system is fully shutdown to allow the inspection and the resolution of the fault. This is done by recalibrating the compromised sensor and by checking the other plant’s components functioning. After these tasks are completed, the system is restarted, the automated control is re-enabled, and the plant restores its operations under nominal conditions.

- *Scenario 2: false low reading of tank pressure*. A wrong provision of the feedback “Measured water pressure in tank” or “Measure of water pressure in tank” (cf. **Figure 4** mislead the control of AV3. Again, this scenario can be verified due to various types of attacks as (e.g.) a stealth attack with FDI, a covert MitM attack, or even a denial-of-service (DoS) attack which prevents the controllers to have an updated vision on the pressure parameter. In the attempt to correct the reported low pressure, the automated controller closes the valve AV3 allowing less air to escape from the tank (i.e., UCA 1) and eventually making the internal pressure dangerously increase. In parallel, the management of inlet water through AV1 remains unchanged (i.e., UCA 2) but the increasing internal pressure impedes water to flow properly, thus compromising the ejector’s ability to draw air. Upon detecting the abnormal system behavior, the human operator is expected to intervene by disabling the automated controls, thereby manually stabilizing the plant in a safe configuration. If executed promptly, this manual control override prevents the situation from escalating and establishes the conditions for a controlled plant shutdown. The operator then proceeds with the shutdown and carries out the necessary restoration procedures (i.e., sensor re-calibration) and checks to bring the plant back to its intended operational state. Once the issue has been resolved, the system is restarted, the automated control is reactivated, and the normal operating conditions are restored, with all process variables returning to their nominal values.

For clarity, the two scenarios will be referred respectively to as Scenario 1 and Scenario 2 throughout the remainder of the paper.

### 3.2.7. Evaluate the system resilience performance

The evaluation of the system's resilience performance was supported by an experimental campaign aimed at collecting operators' behavioral data within the mock-up plant environment during the management of Scenario 1 and Scenario 2. Both expert and novice operators were involved in the simulations to capture a diverse range of human responses to the abnormal system conditions. Specifically, participants with more than one year of experience working at the plant were classified as expert operators, whereas those with less than one year of experience were considered novices. In a human-in-the-loop perspective, all the participants interacted with the plant using a combination of digital control panels and remote interfaces (i.e., desktop and mobile apps), and the plant's physical equipment (i.e., physical sensors and actuators). In each simulation session, the operators' actions were tracked using the motion capture system, generating behavioral data as outlined in **Section 3.2.4**. These temporal sequences were subsequently analyzed and manually related to either the detection or the restoration phase (i.e., from $t_1$ to $t_2$, or from $t_3$ to $t_4$, cf. **Figure 5**).

During the experimental campaign, the sensors equipped on the plant continuously recorded the physical process data, returning a data table as the one described in **Section 3.2.4**. In cases where some dangerous condition was being reached, the system automatically triggered a shutdown, stopping the real plant operations. In such instances, the digital simulation model was used to complete the remaining portions of the time series, ensuring a consistent dataset for the analysis.

Given these two types of data (i.e., the human behavioral data, and the plant performance data) a MATLAB script was developed to assess the whole CSTS resilience under varying operational conditions. Specifically, the script performed Monte Carlo simulations to compute the resilience indicators $R_i$ distributions with respect to different human operator behaviors, drawn from the collected experimental data. The Monte Carlo simulation counted 500 iteration, being set conservatively guaranteeing a <5% error threshold (Driels & Shin, 2004).

At each iteration, the script randomly picks up two human-related data tables, one for the detection activity, and one for the restoration activity. Then, the script draws randomly a plant operational performance curve, among the ones generated during the experimental campaign. The three inputs are merged considering the phases in which the human performance affects the plant's one (see **Section 3.2.5**): the detection time (i.e., the difference between $t_2$ and $t_1$, cf. **Figure 5**) and restoration time (i.e., the difference between $t_4$ and $t_3$, cf. **Figure 5**) are computed by summing up the last column of the detection and restoration temporal sequences, respectively. The two values are then used to truncate the "under attack", "shutdown", and "plant off" segment of the disrupted performance curve ensuring continuity and consistency between data leveraging again the plant's digital model. Specifically, the detection time identifies the time moment in which the attack is resolved and consequently the maximum performance degradation (i.e., $t_2$ and $P_d$, cf. **Figure 5**). This latter is then used to initialize the "shutdown" phase. The restoration time, on the other hand, governs the duration of the "plant off" phase reflecting how long it takes for the operator to perform diagnostics, recalibrations, and prepare the system for its restart. Accordingly, each simulation iteration returns a three-dimensional data object containing: (i) the detection time value; (ii) the restoration time value; and (iii) a table including the plant parameter curves generated during that specific simulation run.

The data generated through the simulation process is then used to compute the resilience indicators $R_i$. To this end, a reference dataset representing the nominal performance of the plant was created by recording approximately two hours of stable operations, during which no disruption was introduced. Any transient behavior was excluded to ensure the dataset reflected only the steady-state system conditions. However, before comparing this reference time series with the disrupted performance curves, a set of adjustments were applied to ensure the most accurate and meaningful computation of system resilience.

The first post-processing step involved aligning the lengths of the two timeseries. Since the disrupted performance curves varied in duration depending on the simulated operator's behavior (i.e., different detection and restoration times), the nominal performance dataset was cyclically trimmed to match the exact length of each disrupted curve. This ensured that the area under the standard performance curve $As_i$ and the area under the disrupted performance curve $Ad_i$ referred to the same temporal extent. Without this step, excessively long standard curves could artificially inflate $As_i$, pushing the resilience score $R_i$ closer to 1, regardless of the actual extent of the system degradation.

The second adjustment addressed the temporal alignment of the two signals. Simply trimming both time series to the same length does not guarantee that periodic behaviors, such as (e.g.) oscillations around a given set-point, are phase-aligned. In other words, even if the two curves may share the same frequency, any time shift between them can result in a discrepancy in the point-by-point area comparison. This misalignment is particularly problematic when considering the absolute difference $|As_i - Ad_i|$ in **Eq. (1)**, as it can lead to a significant underestimation of resilience even when the overall system behavior remains basically the same. To mitigate this issue, the assessment incorporated the Dynamic Time Warping (DTW), a technique designed to align time series that may differ in speed or timing (Müller, 2007). DTW works by identifying the optimal non-linear alignment between two sequences, minimizing the cumulative distance between their elements. Specifically, given two hypothetic time series:

$$\boldsymbol{x} = (x_1, x_2, \dots, x_a, \dots, x_A) \tag{2}$$

and:

$$\boldsymbol{y} = (y_1, y_2, \dots, y_b, \dots, y_B) \tag{3}$$

the DTW constructs a cost matrix $\boldsymbol{D}$ in which each element $D(a, b)$ represents the distance between the $a$-th element of $\boldsymbol{x}$, and the $b$-th element of $\boldsymbol{y}$ calculated as the Euclidean distance between the two:

$$D(a, b) = (x_a - y_b)^2 \tag{4}$$

The goal of the procedure is to find a warping path $\boldsymbol{w}$ within $\boldsymbol{D}$:

$$\boldsymbol{w} = (w_1, w_2, \dots, w_c, \dots, w_C) \tag{5}$$

such as each $w_c = (a_c\ , b_c)$ aligns elements from $\boldsymbol{x}$ and $\boldsymbol{y}$ minimizing the cost function:

$$DTW(\boldsymbol{x}, \boldsymbol{y}) = min \sum_{c\,=\,1}^{C} D(a_c, b_c) \tag{6}$$

where $c = 1, \dots, C$ iterates over all the couples $a$ and $b$ within the warping path $\boldsymbol{w}$.

As a constraint, the $DTW$ path must satisfy: (i) the boundary conditions (i.e., start and end must be at the first and last elements), (ii) the continuity (i.e., steps of size 1), and (iii) the monotonicity (i.e., preserve the time order) of the timeseries $\boldsymbol{x}$ and $\boldsymbol{y}$.

A final refinement was meant to exclude non-anomalous segments from the resilience analysis, ensuring that the resilience metric would not be artificially underestimated simply because it tended to 1 due to large values of both $As_i$ and $Ad_i$. Indeed, to avoid such phenomenon, the resilience assessment was focused specifically on the periods where a meaningful deviation in performance occurs via the implementation of a pragmatic anomaly detection procedure. This latter was empirically tuned and operates by identifying as anomalous any time segment where the disrupted curve deviates by at least 5% from the nominal curve, and this deviation is kept for at least 50 consecutive timesteps. On the other hand, an anomalous period is considered concluded only after 500 consecutive points fall below the 5% deviation threshold. All non-anomalous time segments were excluded from the resilience calculation by setting their values to zero before computing the area under the curve.

For clarity, **Figure 6** presents the standard and disrupted time series of the component S6 (i.e., the water level in tank sensor) both before and after the application of the data post-processing procedures described above.

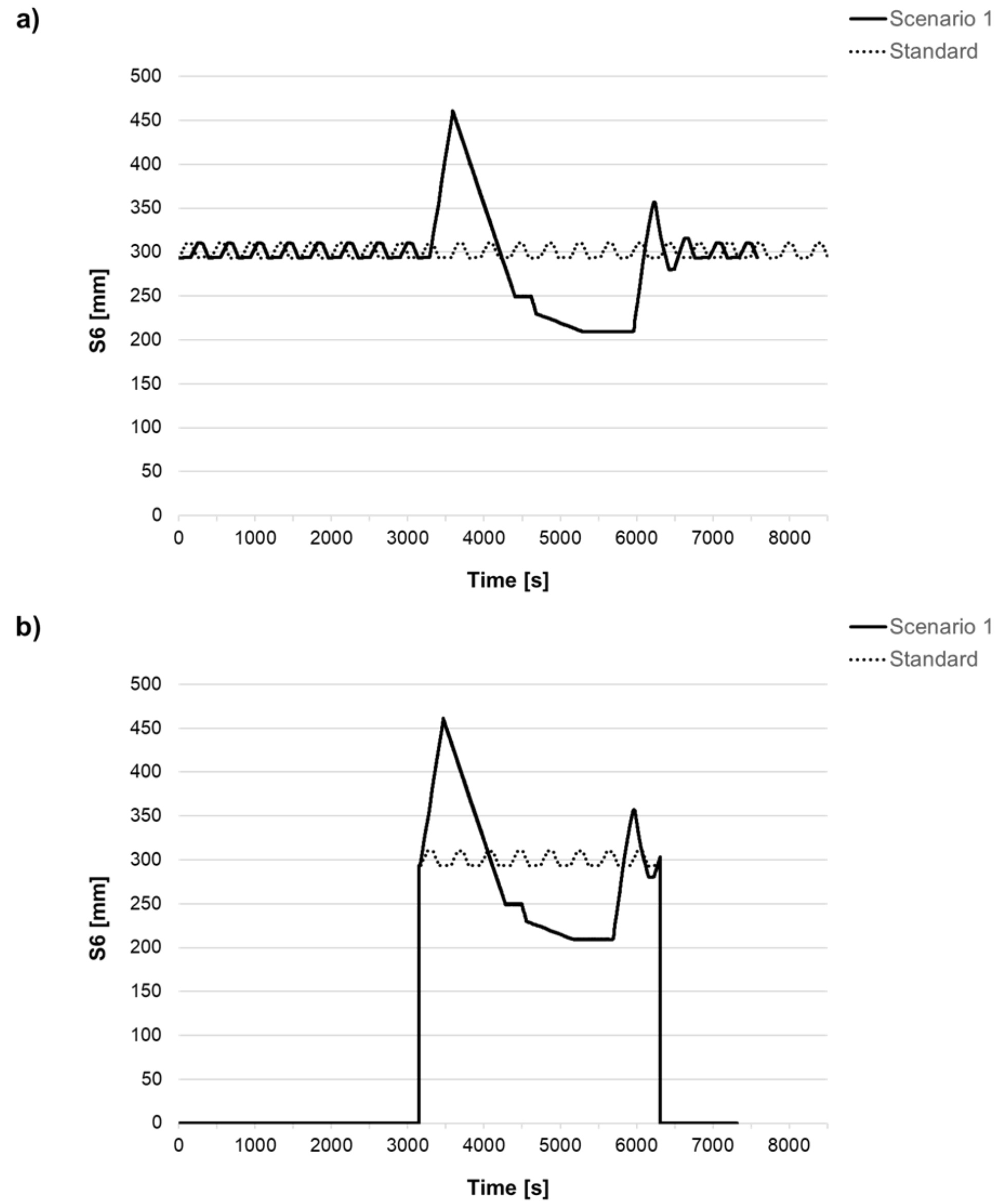

**Figure 6**. Sensor S6 standard and disrupted (Scenario 1) timeseries before (a) and after (b) the post-processing procedure performed before the resilience calculation.

## 4. Simulation results and HHIL resilience assessment

The HHIL simulation runs provided valuable insights into the resilience performance of the CSTS, with a particular focus on the different behaviors of both the expert and the novice operators. This section presents the results obtained for the two simulation scenarios that are compared from two complementary perspectives. The first subsection examines the distribution of: (i) the detection time, (ii) the restoration time, and (iii) the resilience indicators defined in **Section 3.2.5**. The second subsection adopts a broader perspective, introducing two overall resilience metrics: one related to the amount of water produced by the process, and the other related to the volume of air drawn in. Indeed, it is important to recall that the system under analysis is an experimental plant simulating an extraction process, in which water and air represent oil and natural gas, respectively. As such, these two metrics are directly linked to the overall efficiency of the process being simulated.

### 4.1. Results distributions

To assess the statistical significance of the differences observed in results concerning experts and novices, a hypothesis testing approach was adopted. The null hypothesis ($H_0$) stated that "the resilience of expert and novice operators is equal", while the alternative hypothesis ($H_1$) stated that "the resilience of expert and novice operators is different". Each resilience indicator was first tested for normality using the Shapiro-Wilk test. Since no indicator yielded a p-value greater than 0.05, the assumption of normality was rejected. Consequently, the non-parametric Mann-Whitney U test was applied to compare each pair of resilience indicators between novice and expert operators across both scenarios. The obtained p-values are reported in **Table 2**, data confirming the $H_0$ hypothesis (p-value > 0.05), thus showing no significance difference between the two distributions, are underlined in **Table 2**.

**Table 2.** Mann-Whitney p-value results for resilience variables in the two scenarios. Underlined values represent no significant difference between the expert and novice distributions.

| Variable | p-value (Mann-Whitney) for Scenario 1 | p-value (Mann-Whitney) for Scenario 2 |
|---|---|---|
| Detection time | $1.254 \times 10^{-120}$ | $1.515 \times 10^{-93}$ |
| Restoration time | $\underline{2.494 \times 10^{-1}}$ | $\underline{4.206 \times 10^{-1}}$ |
| $R_{S1}$ | $5.838 \times 10^{-15}$ | $1.767 \times 10^{-5}$ |
| $R_{S2}$ | $2.071 \times 10^{-2}$ | $\underline{3.833 \times 10^{-1}}$ |
| $R_{S5}$ | $9.264 \times 10^{-19}$ | $\underline{6.093 \times 10^{-2}}$ |
| $R_{S6}$ | $8.387 \times 10^{-111}$ | $\underline{9.833 \times 10^{-1}}$ |
| $R_{S7}$ | $8.621 \times 10^{-82}$ | $1.353 \times 10^{-7}$ |
| $R_{AV1}$ | $6.088 \times 10^{-48}$ | $8.876 \times 10^{-6}$ |
| $R_{AV2}$ | $1.563 \times 10^{-31}$ | $2.560 \times 10^{-2}$ |
| $R_{AV3}$ | $8.521 \times 10^{-5}$ | $2.189 \times 10^{-15}$ |

On this premises, **Figure 7** and **Figure 8** include the box plots for the resilience indicators related to the eight plant's component (cf. **Table 1**) in Scenario 1 and Scenario 2, respectively. Additionally each figure details the difference between the expert (cf. **Figure 7a** and **Figure 8a**) and novice (cf. **Figure 7b** and **Figure 8b**) operators' distributions. For completion, numerical values (mean, standard deviation, minimum value, maximum value, and 25%, 50%, and 70% percentiles) are included in the **Table 3**, **Table 4**, **Table 5**, and **Table 6**.

In the first scenario, the cyber-attack compromises the sensor S6, responsible for monitoring the tank's water level. The expert operators identify the failure significantly faster than novices, with an average detection time of 261.798 seconds against 433.982 seconds: a difference of nearly three minutes. This gap points at the expected experts' superior performance, and to their deeper understanding of the plant's normal behavior. Accordingly, for them it was much easier to capture the difference in the plant's functioning by means of (e.g.) subtle noise when the disruption occurred. Moreover, the experts exhibit overall a much lower variability in detection times, with a standard deviation of 56.155 seconds versus 105.556 seconds for novices, indicating, overall, a more consistent and effective response. In contrast, looking at the restoration times, they are nearly identical between the two groups, on average: 30.716 seconds for experts against the 30.612 seconds for novices, both with minimal variation (1.654 versus 1.679), as expected from the p-value in **Table 2**. This result may suggest that once the fault is recognized, the recovery procedures are well-defined and equally accessible to both groups, which basically act the same.

The resilience indicators related to the plant's technical components offer further distinctions. If looking at the compromised sensor S6, expert operators achieve a significantly higher resilience score (0.818 versus 0.706), along with a lower variability (0.039 versus 0.074), pointing at a more reliable capacity to stabilize the system despite losing direct visibility on the water level. This may indicate that the experts are likely compensating the missing information by interpreting indirect signals or inferring other system states. For sensors S1, S2, and S5, resilience values are broadly similar. Novices slightly outperform experts in S1, but the difference remains negligible (0.883 versus 0.877), and the variability remains low for both (0.015 versus 0.014). Regarding S2, all operators fail in maintaining an adequate input water flow rate, with average resilience values as similar as low (0.375 versus 0.376). In S5 instead, both groups show an equally high average resilience (0.881 for both), suggesting an effective pressure management even under the attack on S6. The three indicators on sensors S1, S2, S3 show a shared intention among operator to prioritize the pressure management, which is more critical, instead than the flow one. Experts also show a slightly better performance in S7 (0.755 versus 0.732), with a bit lower variability (0.013 versus 0.018), indicating better control over the air intake. The resilience indicators related to the valve controls reveal some additional strategic differences between the two types of users. The experts exhibit a higher resilience in the behavior of AV1 with an average of 0.334 compared to 0.281 for novices, with basically equal variability (0.051 versus 0.052). This may reinforce the idea that expert operators place a stronger effort on stabilizing the input when the direct feedback on the tank level is compromised. They also outperform novices in the AV2 and AV3 controls, which handle the air and water discharge. Although AV2 shows a really low resilience, experts attain a higher mean value (0.084 versus 0.049) and a slightly greater variability (0.048 versus 0.037), implying a potentially riskier approach. In AV3, experts again show slightly higher average resilience (0.487 versus 0.480) with a bit reduced variability (0.021 versus 0.026), reflecting an albeit minimal steadier control.

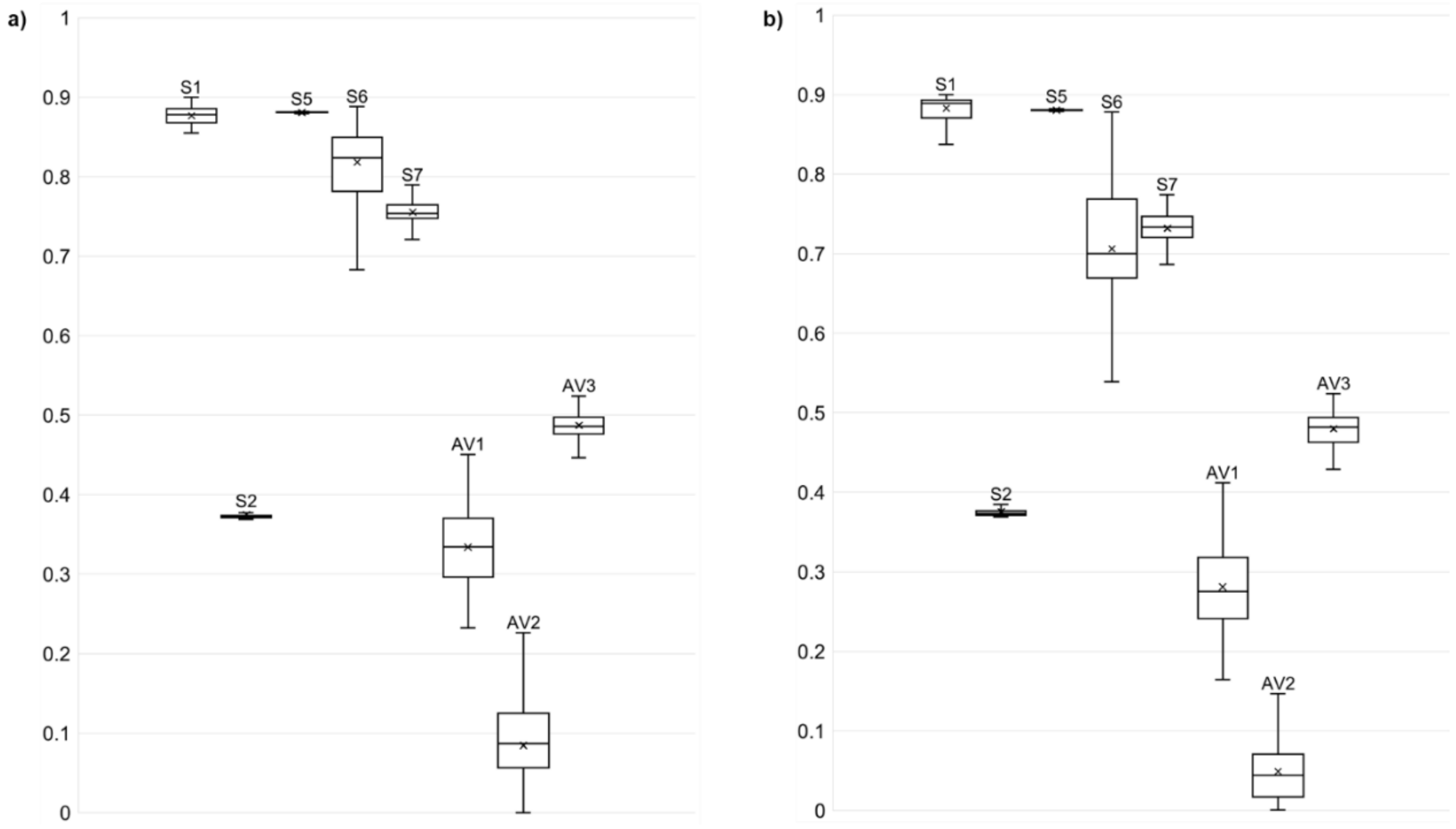

**Figure 7**. Distribution of resilience indicators per plant's components in Scenario 1 grouped by expert (a) and novice (b) operators.

In the second scenario, the cyber-attack targets S5, which is instrumental to monitor the pressure inside the tank, a central variable for ensuring system safety. Again, on average expert operators detect the failure more quickly (238.170 seconds versus 288.622 seconds), almost one minute faster than novices, reinforcing their higher situation awareness. Restoration times remain not significantly different between the groups (cf. **Table 2**), indicating a possible consistent application of the recovery protocols even in Scenario 2. Notably, when looking at the technical-related resilience indicators, the difference between the two is in this case much less evident, with expert operators displaying even a slightly lower resilience in some components (i.e., S1 and S7), all of which are closely linked to the pressure management in both the pipes and the tank. This may suggest a more aggressive control strategy, where rapid interventions introduce temporary instabilities that ripple through the system. Despite this, both groups demonstrate high average resilience in S5 (0.803 for both), reflecting an effective system recovery strategy following the disruption on tank pressure. On the contrary, both the groups demonstrate an extremely low average resilience value for S1 (0.246 versus 0.249) which was among the best performing in Scenario 1. Moreover, also in this case, the input flow rate remains the most disrupted performance metric overall (on average, 0.059 for both), underlining the widespread impact of the pressure sensor compromise up to the earlier stages of the system. Valve control related results once again indicate that experts prioritize input stability, with slightly higher resilience in AV1 (0.241 versus 0.228). Conversely, they exhibit a marginally lower resilience in AV2 (0.212 versus 215) and AV3 (0.342 versus 3.64), suggesting an albeit minimal trade-off consisting in sacrificing the tank stability – which was unsensed because of the cyber-attack – to regain control over the input pressure parameter. However, the three resilience indicators related to the valves control remains among the lower, in a way similar to Scenario 1.

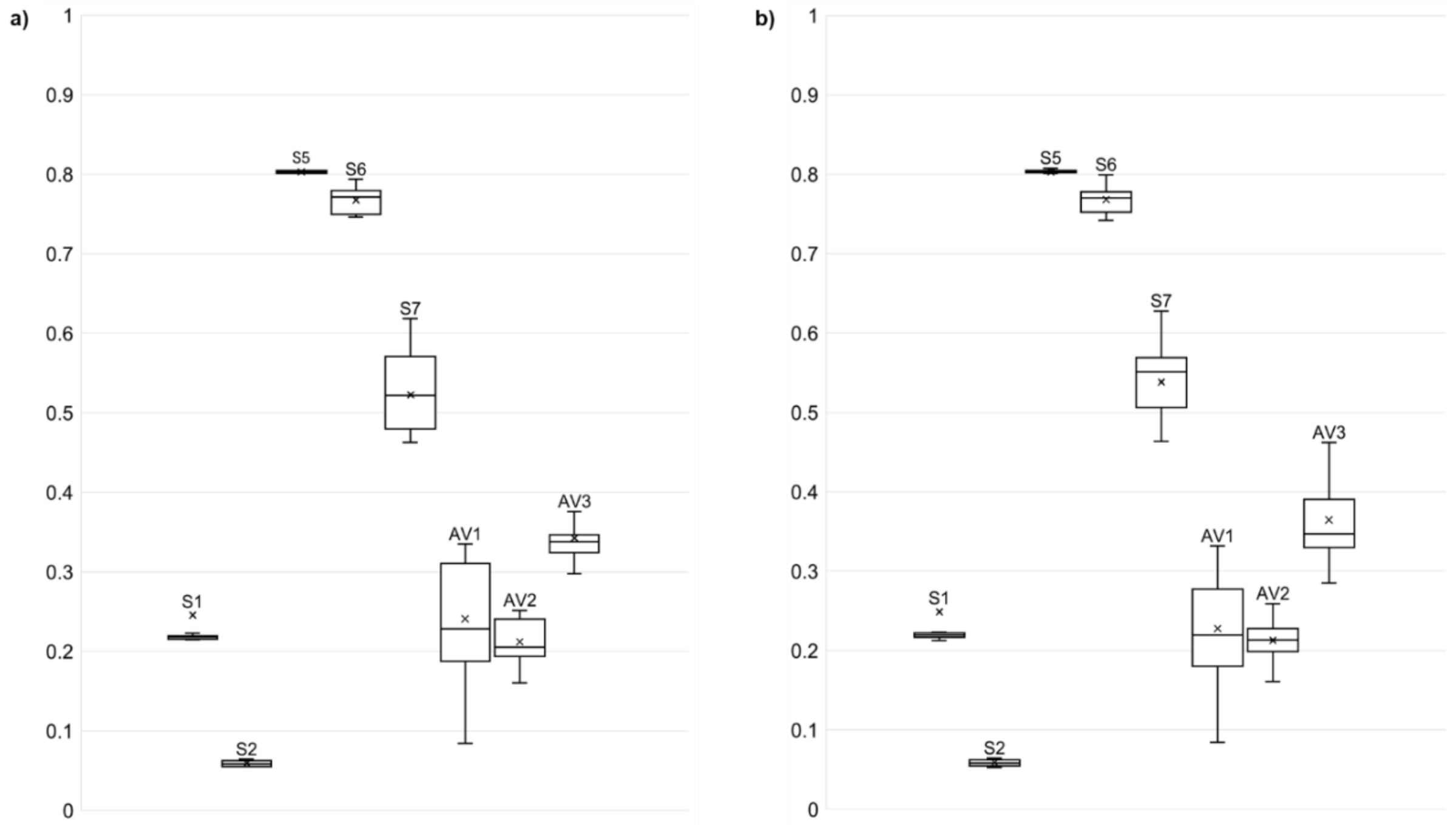

**Figure 8**. Distribution of resilience indicators per plant’s components in Scenario 2 grouped by expert (a) and novice (b) operators.

**Table 3.** Numerical results for the distributions of detection time, restoration time, and resilience indicators in Scenario 1 for the expert operator.

| | Detection time [s] | Restoration time [s] | $R_{S1}$ | $R_{S2}$ | $R_{S5}$ | $R_{S6}$ | $R_{S7}$ | $R_{AV1}$ | $R_{AV2}$ | $R_{AV3}$ |
|---|---|---|---|---|---|---|---|---|---|---|
| **Mean** | 261.798 | 30.716 | 0.877 | 0.375 | 0.881 | 0.818 | 0.755 | 0.334 | 0.084 | 0.487 |
| **Standard deviation** | 56.155 | 1.654 | 0.015 | 0.007 | 0.001 | 0.039 | 0.013 | 0.051 | 0.048 | 0.021 |
| **Minimum** | 143.000 | 28.000 | 0.776 | 0.369 | 0.873 | 0.675 | 0.720 | 0.232 | 0.001 | 0.429 |
| **25%** | 220.000 | 29.000 | 0.868 | 0.371 | 0.881 | 0.782 | 0.747 | 0.296 | 0.057 | 0.476 |
| **50%** | 254.000 | 30.000 | 0.878 | 0.373 | 0.881 | 0.824 | 0.754 | 0.334 | 0.087 | 0.486 |
| **75%** | 301.000 | 32.000 | 0.886 | 0.374 | 0.881 | 0.849 | 0.765 | 0.370 | 0.125 | 0.498 |
| **Maximum** | 417.000 | 33.000 | 0.900 | 0.393 | 0.882 | 0.888 | 0.798 | 0.451 | 0.226 | 0.577 |

**Table 4.** Numerical results for the distributions of detection time, restoration time, and resilience indicators in Scenario 1 for the novice operator.

| | Detection time [s] | Restoration time [s] | $R_{S1}$ | $R_{S2}$ | $R_{S5}$ | $R_{S6}$ | $R_{S7}$ | $R_{AV1}$ | $R_{AV2}$ | $R_{AV3}$ |
|---|---|---|---|---|---|---|---|---|---|---|
| **Mean** | 433.982 | 30.612 | 0.883 | 0.376 | 0.881 | 0.706 | 0.732 | 0.281 | 0.049 | 0.480 |
| **Standard deviation** | 105.556 | 1.679 | 0.014 | 0.007 | 0.001 | 0.074 | 0.018 | 0.052 | 0.037 | 0.026 |
| **Minimum** | 184.000 | 28.000 | 0.820 | 0.369 | 0.879 | 0.481 | 0.687 | 0.165 | 0.001 | 0.429 |
| **25%** | 358.500 | 29.000 | 0.870 | 0.371 | 0.880 | 0.670 | 0.720 | 0.242 | 0.017 | 0.463 |
| **50%** | 426.500 | 30.000 | 0.889 | 0.373 | 0.881 | 0.700 | 0.733 | 0.275 | 0.044 | 0.482 |
| **75%** | 509.000 | 32.000 | 0.893 | 0.377 | 0.881 | 0.768 | 0.747 | 0.318 | 0.071 | 0.494 |
| **Maximum** | 709.000 | 33.000 | 0.900 | 0.393 | 0.882 | 0.878 | 0.774 | 0.446 | 0.190 | 0.524 |

**Table 5.** Numerical results for the distributions of detection time, restoration time, and resilience indicators in Scenario 2 for the expert operator.

| | Detection time [s] | Restoration time [s] | $R_{S1}$ | $R_{S2}$ | $R_{S5}$ | $R_{S6}$ | $R_{S7}$ | $R_{AV1}$ | $R_{AV2}$ | $R_{AV3}$ |
|---|---|---|---|---|---|---|---|---|---|---|
| **Mean** | 238.170 | 1590.710 | 0.246 | 0.059 | 0.803 | 0.768 | 0.523 | 0.241 | 0.212 | 0.342 |
| **Standard deviation** | 18.253 | 118.542 | 0.065 | 0.003 | 0.001 | 0.015 | 0.043 | 0.056 | 0.024 | 0.036 |
| **Minimum** | 193.000 | 1429.000 | 0.215 | 0.055 | 0.801 | 0.746 | 0.463 | 0.084 | 0.161 | 0.279 |
| **25%** | 225.750 | 1442.000 | 0.215 | 0.055 | 0.801 | 0.750 | 0.480 | 0.188 | 0.194 | 0.324 |
| **50%** | 240.000 | 1606.000 | 0.218 | 0.058 | 0.803 | 0.771 | 0.522 | 0.228 | 0.205 | 0.338 |
| **75%** | 252.000 | 1737.000 | 0.220 | 0.063 | 0.804 | 0.779 | 0.571 | 0.310 | 0.241 | 0.346 |
| **Maximum** | 280.000 | 1739.000 | 0.405 | 0.064 | 0.805 | 0.793 | 0.618 | 0.335 | 0.252 | 0.460 |

**Table 6.** Numerical results for the distributions of detection time, restoration time, and resilience indicators in Scenario 2 for the novice operator.

| | **Detection time [s]** | **Restoration time [s]** | **$R_{S1}$** | **$R_{S2}$** | **$R_{S5}$** | **$R_{S6}$** | **$R_{S7}$** | **$R_{AV1}$** | **$R_{AV2}$** | **$R_{AV3}$** |
|---|---|---|---|---|---|---|---|---|---|---|
| **Mean** | 288.622 | 1597.218 | 0.249 | 0.059 | 0.803 | 0.768 | 0.538 | 0.228 | 0.215 | 0.364 |
| **Standard deviation** | 40.528 | 120.752 | 0.062 | 0.004 | 0.001 | 0.016 | 0.037 | 0.054 | 0.019 | 0.046 |
| **Minimum** | 207.000 | 1429.000 | 0.213 | 0.053 | 0.801 | 0.742 | 0.463 | 0.084 | 0.161 | 0.285 |
| **25%** | 257.000 | 1442.000 | 0.216 | 0.055 | 0.802 | 0.752 | 0.506 | 0.181 | 0.199 | 0.330 |
| **50%** | 283.000 | 1606.000 | 0.219 | 0.058 | 0.803 | 0.770 | 0.551 | 0.219 | 0.213 | 0.347 |
| **75%** | 315.000 | 1737.000 | 0.223 | 0.063 | 0.804 | 0.778 | 0.569 | 0.277 | 0.228 | 0.390 |
| **Maximum** | 392.000 | 1739.000 | 0.457 | 0.064 | 0.808 | 0.799 | 0.628 | 0.332 | 0.259 | 0.462 |

## 4.2. Overall resilience performance

As mentioned above, this paragraph takes a broader perspective on the CSTS resilience by means of: (i) the output water flow rate, which relates to the system's ability to sustain a consistent water (that in the experimental layout mimics the actual plant's oil output) ; and (ii) the input air flow rate, which relates to the system's capacity to maintain the suction efficiency at the ejector inlet, eventually ensuring a stable pressure support for the reservoir.

Again, the two dimensions were checked for consistency performing the Shapiro-Wilk test at first. None of the dimensions passed the test (p-value < 0.05) thus the Mann-Whitney U test was conducted. Similarly to **Table 2**, the obtained results are presented in **Table 7**.

**Table 7**. Mann-Whitney p-value results for resilience related to the output water flow rate, and resilience related to the input air flow rate in the two scenarios.

| Variable | p-value (Mann-Whitney) for Scenario 1 | p-value (Mann-Whitney) for Scenario 2 |
|---|---|---|
| Output water flow rate | $6.874 \times 10^{-117}$ | $6.687 \times 10^{-20}$ |
| Input air flow rate | $8.621 \times 10^{-82}$ | $1.353 \times 10^{-7}$ |

**Figure 9** and **Figure 10** include two scatterplots related to the aforementioned metrics for Scenario 1, where the resilience measures are plotted against the operators' detection time. Only the detection time was considered as the results in **Section 4.1** shows how the variations in the restoration time are not significant. Both plots clearly demonstrate that rapid anomaly detection is the most critical factor driving system resilience.

In the water flow plot (cf. **Figure 9**), the overall performance loss remains substantially high for all operators. Experts are clustered within a detection window of 150 to 400 seconds, with an inter-quartile range (IQR) equal to 81 seconds, and resilience values largely falling between 0.40 and 0.55. This indicates their ability to identify the issue promptly, and their consistency in controlling the water output, regardless of any variation in detection time. This consistency is further confirmed by a low IQR in the resilience score (i.e., equal to 0.027). In contrast, the novice operators exhibit a much wider distribution, with most detection times ranging from 350 to 700 seconds (IQR = 150.5 seconds), and even lower resilience scores, generally between 0.15 and 0.45. Their resilience variability is also slightly higher, as indicated by a larger IQR (i.e., equal to 0.139), highlighting a less consistent control over the water output. This supports the idea that the detection delay is a major driver for recovery effectiveness. However –several data points from novice operators show a sort of rebound in resilience for detection times beyond 600 seconds, reaching levels comparable to those associated with much earlier detections. While this apparent recovery may not directly reflect improved operators' skills, it remains a notable result. One possible explanation is that some novices, when faced with ambiguous signals or minor anomalies, adopt a passive "let-it-run" approach, delaying the intervention out of caution. During this period of inaction, given that the cyber-attack targets the level sensor, only, the plant's PID controllers continue to regulate pressures and flows, keeping the system relatively stable until corrective actions are finally taken. Nevertheless, a more intriguing interpretation is that less experienced operators may, paradoxically, benefit from a less constrained and more improvisational management approach. They might be forced to adapt without relying on established routines or prior expectations, employing a more flexible and exploratory strategy that, under stress, results in unexpectedly better performance. This result suggests how extensive experience – though generally advantageous – could sometimes lead to more rigid, less adaptive behaviors that ultimately limit system's resilience itself. This could be argued also with the experts' resilience plots, which appear more normally distributed.

The air input flow rate scatterplot in **Figure 10** reinforces the broader conclusions commented above, but with a less evident rebound distortion, and much higher resilience values. Experts resilience scores range between 0.75 and 0.80 (IQR = 0.017), reflecting an overall good management of the suction performance. In contrast, the novices show a greater resilience decline as detection delays increase, dropping down to a resilience value of 0.70 or lower, for detection times beyond 500 seconds. However, their variability remains minimal, as demonstrated by the low IQR in the resilience distribution (i.e., equal to 0.026).

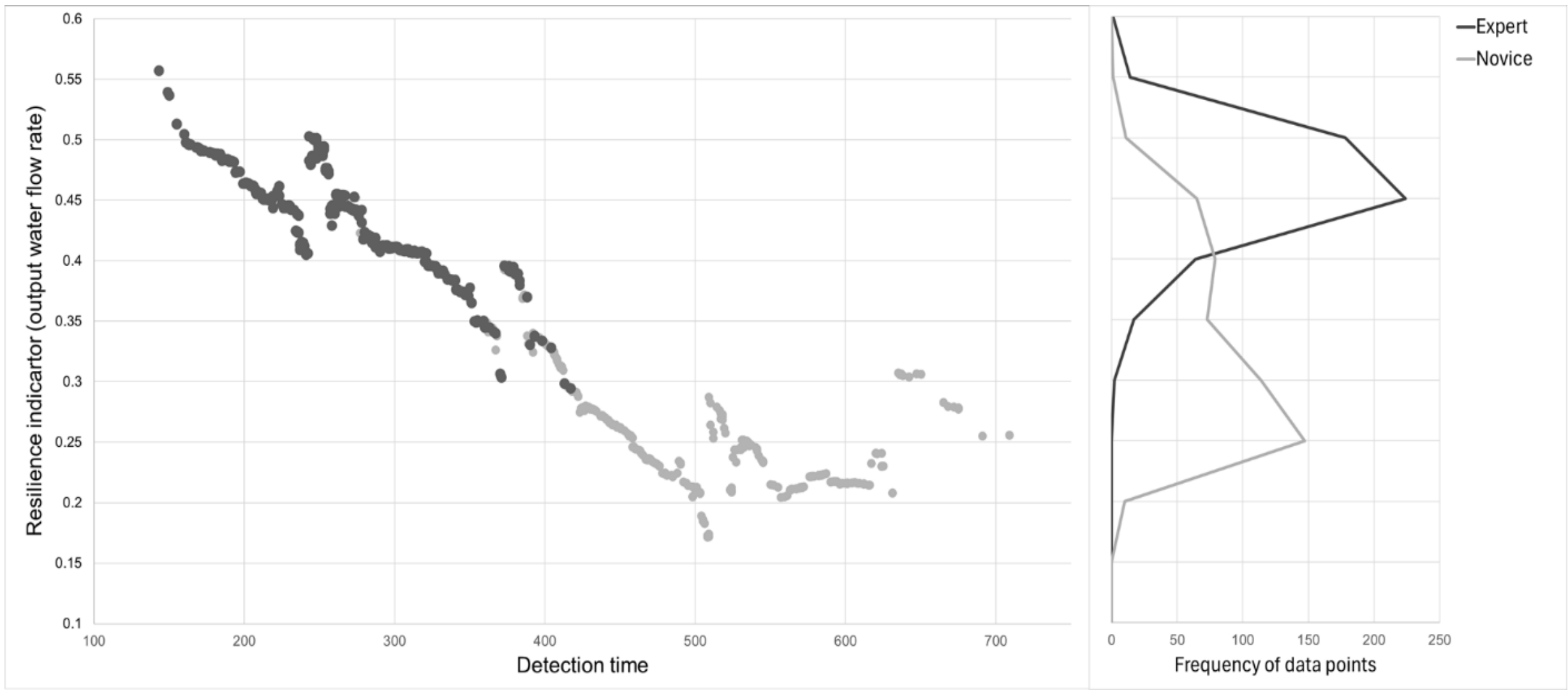

**Figure 9**. Overall process performance resilience metric (i.e., output water flow rate) for Scenario 1 distributed by detection time for expert (dark grey) and novice (light grey) operators with frequency of occurrence on the right. In the frequency plots, the count of points within each range is represented at the upper bound of the range.

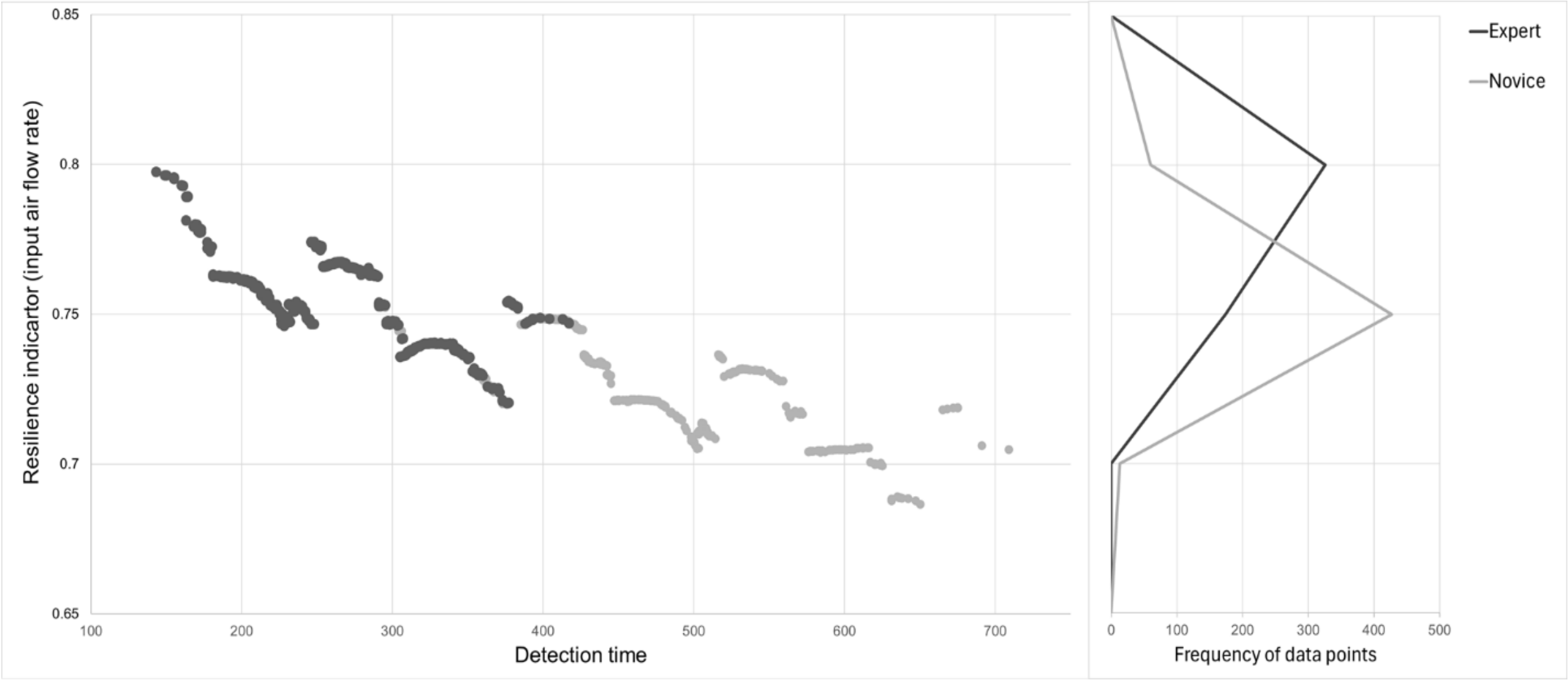

**Figure 10**. Overall process performance resilience metric (i.e., input air flow rate) for Scenario 1 distributed by detection time for expert (dark grey) and novice (light grey) operators with frequency of occurrence on the right. In the frequency plots, the count of points within each range is represented at the upper bound of the range.

Similarly, **Figure 11** and **Figure 12** shows the scatter plots of the detection time against the resilience computed by means of the water output flow rate and the air input flow rate, for Scenario 2, in which the tank-pressure sensor S5 is compromised. As for Scenario 1, differences in the restoration times were not significant, and only the detection time was considered in the analysis.

In both plots, expert operators are tightly clustered within a detection window of 200 to 350 seconds (IQR = 26 seconds), achieving a moderate yet stable resilience performance. Their resilience values range approximately between 0.50 and 0.65 for both the water output and the air input flow rates. The distributions of the two resilience dimensions show minimal variabilities, with IQR = 0.053 for the output water flow rate, and IQR = 0.091 for the air inlet. Interestingly, novice operators exhibit a similar overall pattern, with resilience indicators stabilizing within a comparable range of 0.45 to 0.60, and overall low IQRs (i.e., 0.017 for the water output, and 0.063 for the air inlet), yet demonstrating a wider variability in detection time (IQR = 59.25 seconds). However, a notable divergence appears in the water output flow rate where around 350 seconds, their performance shows a clear decline, stabilizing at a lower average of approximately 0.50.

This threshold-driven drop in performance may reflect an important characteristic of the pressure-sensitive dynamics investigated in Scenario 2. As pointed out in the previous findings, the early detection proves to be an even critical aspect in this case. Once a sensor breach goes undetected beyond a certain point, the pressure oscillations intensify, preventing both a manual adjustment, and the PID-based stabilization. This leads to the identification of a kind of “non-returning point” (i.e., when detection goes over 350 seconds), to be avoided at all costs, where the system control becomes significantly more difficult. In this context, the passive “let-it-run” approach that may have benefited novices in Scenario 1 appears counterproductive in Scenario 2. The pressure-based disruption brings the system to move more, and more permanently from its target values, making it increasingly difficult to recover and sustain high resilience performance if the intervention is too delayed.

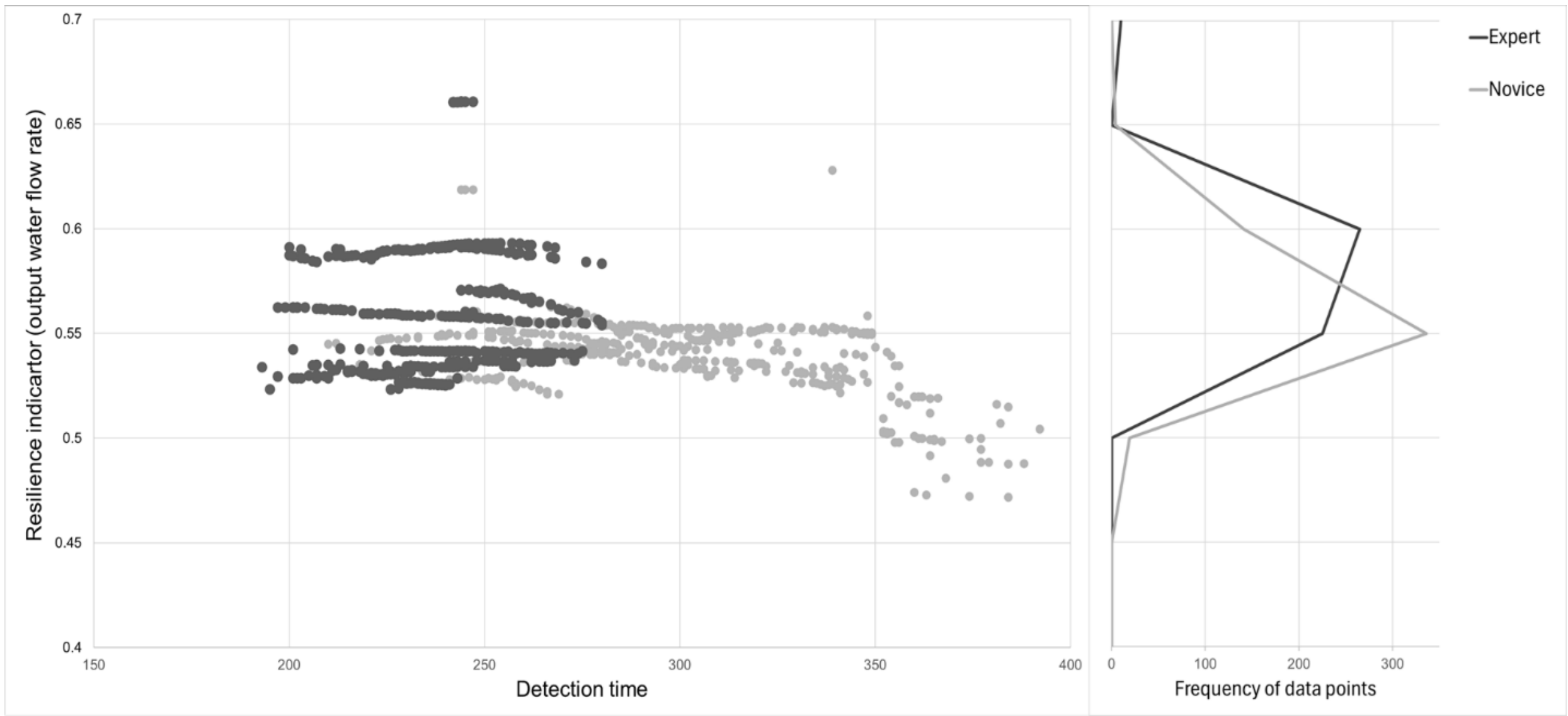

**Figure 11**. Overall process performance resilience metric (i.e., output water flow rate) for Scenario 2 distributed by detection time for expert (dark grey) and novice (light grey) operators with frequency of occurrence on the right. In the frequency plots, the count of points within each range is represented at the upper bound of the range.

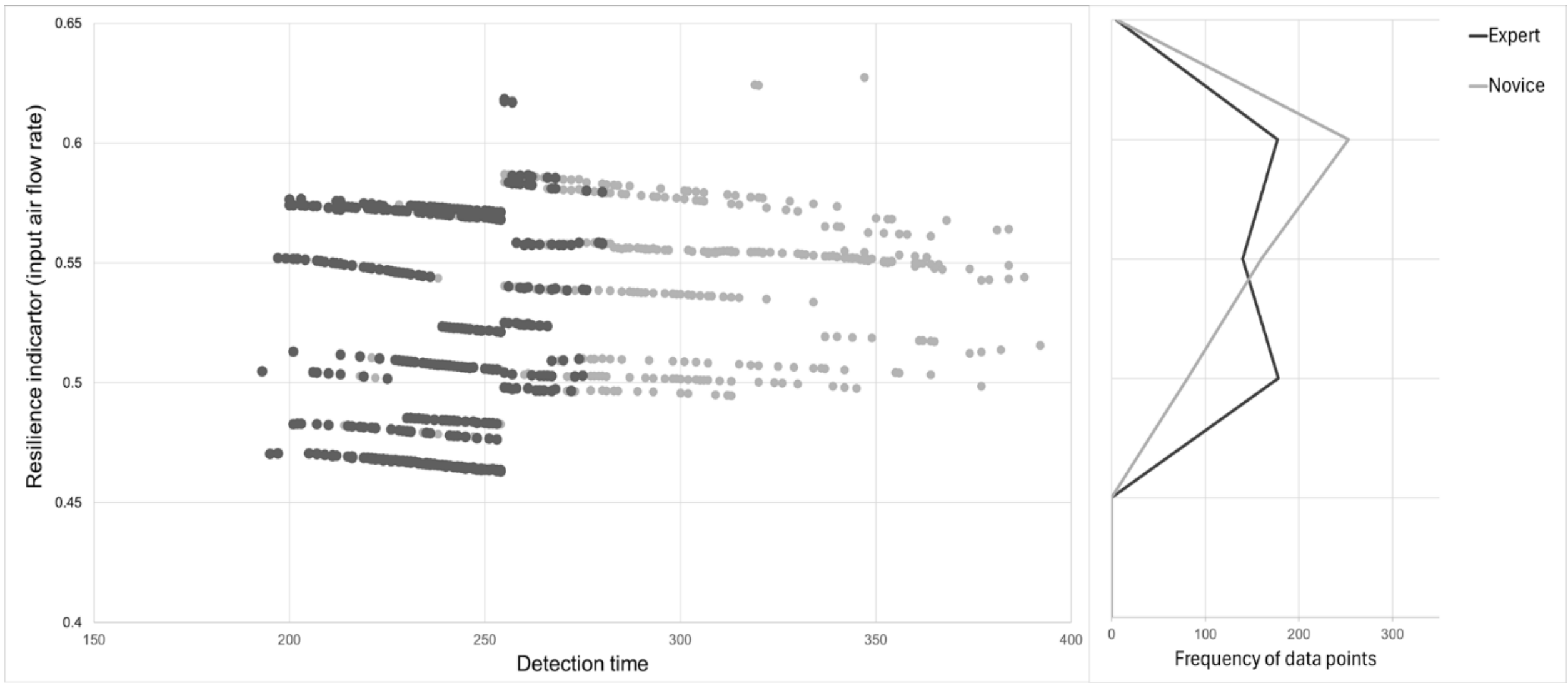

**Figure 12**. Overall process performance resilience metric (i.e., input air flow rate) for Scenario 2 distributed by detection time for expert (dark grey) and novice (light grey) operators with frequency of occurrence on the right. In the frequency plots, the count of points within each range is represented at the upper bound of the range.

## 5. Discussion

Although the entire set of experiments was conducted on a simulated, non-critical CSTS, this section aims to explore how the results of the resilience analysis can inform decision-makers and system managers by offering practical and actionable insights.

Firstly, it is worth noting that the results are based on assessment inspired by the largely sued concept of “resilience triangle” (cf. **Section 3.2.5**), which has been recently acknowledged to have limited representational power (Eisenberg et al., 2025). While it can mask important dimensions of resilience, particularly those related to adaptation dynamics and recovery pathways, it remains a simple communicable proxy for capturing system performance under disruption like the cyber-attacks reported in this study. As such, the approach has been applied carefully and critically, with an awareness of its limitations. It has been complemented by additional considerations to ensure that the analysis does not reduce resilience to a single geometric abstraction. Rather, the resilience triangle is used here as an accessible entry point, while recognizing that more nuanced approaches, such as multi-dimensional performance envelopes or adaptive capacity metrics, will be necessary to further advance CSTS’ resilience assessment.

In this regard, the comparative analysis across the two simulation scenarios presented in **Section 4** highlights the complexity and the significance of human decision-making while managing a system under a cyber-attack. The simulations confirm that human behavior, particularly in terms of early anomaly detection, is one of the most critical factors for ensuring the overall system resilience. Nevertheless, while expert operators consistently deliver faster and more reliable responses, their strategies tend to be much more rigid. This is effective in many scenarios, but novice operators, though slower and more variable in their responses, occasionally exhibit some surprisingly resilient outcomes (especially in Scenario 1) when adopting a more improvised and flexible approach.

From a managerial perspective, these findings raise important questions about how much “freedom” should be granted to operators, particularly in light of a trade-off between controllability and adaptability described by (Grote et al., 2018). While experience and intensive training generally lead to more reliable and stable performance, they may also constrain operators' capacity to improvise in unforeseen situations. This tension directly echoes Woods' Theory of Graceful Extensibility (Woods, 2018), which emphasizes the importance of sustaining the capacity to adapt as boundaries are approached or exceeded. In this sense, resilience is not solely a matter of maintaining controllability, but of cultivating the potential for units of adaptive behaviors to flexibly extend their performance, thereby preserving effectiveness under surprising conditions.

In this context, building resilience in complex systems likely requires a combination of both deep expertise, the capacity, and the possibility, to deviate from standard routines, when necessary. Training programs are important to foster individual competence, but they alone are insufficient. Reflecting the controllability/resilience duality, systems must be designed to balance reliable control with the ability to adapt when conditions diverge from expectations. Likewise, the system's capacity to extend performance when boundaries are reached cannot rest solely on individual skill. This calls for system designs that deliberately integrate both dimensions: providing operators with structured knowledge and stable routines, while also embedding affordances for flexible coordination, anomaly detection, and escalation. In this sense, resilience is engineered into the socio-technical system.

For example, a promising avenue for improving detection time supporting less experienced operators lies in enhancing the human-machine interface. During various simulations, experts were often able to quickly detect faults by triangulating information from multiple sources, which were not shown on the available interface. These expert-level heuristics could potentially be formalized into clear visual indicators integrated into the operators' interfaces, possibly enhancing the novices' situation awareness by making such tacit knowledge more accessible and explicit.

The simulations also highlight how the failure of a single sensor – particularly S5 – can have a significant impact on the entire system. With respect to the human-machine interface content, this finding suggests the importance of incorporating some kind of measurement redundancy. This could be achieved either through backup sensors or through estimation systems that infer missing parameters using indirect data, thereby improving the availability of critical information on the operator's display. For example, the water level inside the tank could be estimated from the flow rates and the known dimensions of the tank and the pipelines. Accordingly, when a sensor might be identified as unavailable, such inferred values could be automatically fed to the interface, ensuring that such information remains available to the operators.

The widespread impact of cyber-attacks also prompts two further reflections. In its current configuration, the automation manages the system in a point-to-point fashion, lacking a truly systemic perspective, which was peculiar in humans, instead. This highlights the need to rethink the control logic governing the valves: rather than having each valve operating independently to regulate a single parameter, a more holistic control strategy could be implemented to coordinate valve actions in proportion to one another. On the contrary, instead, the system's vulnerability to cascading disruptions may also suggest the potential benefit of segmenting it into functional zones. Accordingly, a modular approach could help localize and contain anomalies before they propagate system-wide. Although implementing such fault isolation strategies may be challenging in small, tightly coupled systems like the one simulated here, the concept remains relevant for larger, more structured industrial environments.

## 6. Conclusion

This paper introduced a HHIL simulation approach to enhance the systemic yet quantitative resilience assessment of CSTS, building upon the STPA-Sec/S methodology. By integrating human behavior and physical process data into a simulation environment, the proposed approach aimed at offering a more realistic representation of complexity in system facing cyber-related disruptions. The approach was tested through an experimental oil and gas plant managed by two distinct operator groups (i.e., experts and novices). The application demonstrated that the HHIL simulations were capable of effectively capturing the systemic interplay between human decision-making and system performance. Overall, the obtained results indicated that expert operators generally responded more quickly and consistently, while novices occasionally displayed adaptive behaviors that led to unexpectedly resilient outcomes. These findings highlight the importance of balancing technical expertise while fostering the adaptive capacity of operators, as both dimensions appear closely linked to ensuring the overall CSTS resilience and thus, graceful extensibility. Being here emerged only as hints, future research shall explore in more detail how these dimensions actually relate to real system resilience. At present, the implementation is limited to custom-designed experimental environments, and integrating it into operational processes remains an open challenge.

In conclusion, we believe the proposed HHIL approach represents a little yet meaningful step forward in bridging qualitative system-theoretic analysis with more quantitative assessments. By providing a more realistic understanding of human-machine interdependencies in complex systems, this approach lays an important foundation for the development of safer and more resilient industrial systems in an increasingly complex and digitalized world.

## Acknowledgement

This research has been funded by the European Union – NextGenerationEU under the National Recovery and Resilience Plan (PNRR) – Mission 4 Education and Research – Component 2 From research to business - Investment 1.1, Prin 2022 Notice announced with DD No. 104 of 2/2/2022, entitled RESIST – RESilience management to Industrial Systems Threats, proposal code 2022YSAE2X – CUP B53D23006650006.

## CRediT roles

*Francesco Simone*: Conceptualization, Data curation, Formal analysis, Investigation, Methodology, Resources, Software, Validation, Visualization, Writing – original draft, Writing – review & editing; *Marco Bortolini*: Data curation, Resources, Software, Writing – review & editing; *Giovanni Mazzuto*: Data curation, Resources, Software, Writing – review & editing; *Giulio Di Gravio*: Supervision, Writing – review & editing; *Riccardo Patriarca*: Conceptualization, Funding acquisition, Investigation, Methodology, Project administration, Supervision, Writing – review & editing.